\documentclass[fleqn,usenatbib]{mnras}

\usepackage{newtxtext,newtxmath}

\usepackage[T1]{fontenc}

\DeclareRobustCommand{\VAN}[3]{#2}
\let\VANthebibliography\thebibliography
\def\thebibliography{\DeclareRobustCommand{\VAN}[3]{##3}\VANthebibliography}

\usepackage{graphicx}	
\usepackage{amsmath} 
\usepackage{pdflscape}
\usepackage{array}

\title[Metallicity of RRLs from {\it Gaia} DR3]{Metallicity of RR Lyrae stars from the {\it Gaia} Data Release 3 catalogue computed with Machine Learning algorithms.}

\author[T. Muraveva et al.]{
Tatiana Muraveva$^{1}$\thanks{E-mail: tatiana.muraveva@inaf.it},
Andrea Giannetti$^{2}$,
Gisella Clementini$^{1}$, 
Alessia Garofalo$^{1}$
and Lorenzo Monti$^{1}$
\\
$^{1}$ INAF - Osservatorio di Astrofisica e Scienza dello Spazio di Bologna, Via Piero Gobetti 93/3, Bologna 40129, Italy\\
$^{2}$ Istituto di Radioastronomia - INAF, Via Piero Gobetti 101, Bologna 40129, Italy\\
}

\date{Accepted XXX. Received YYY; in original form ZZZ}

\pubyear{2015}

\begin{document}
\label{firstpage}
\pagerange{\pageref{firstpage}--\pageref{lastpage}}
\maketitle

\begin{abstract}

We present new $P -\phi_{31}-{\rm [Fe/H]}$ and $P -\phi_{31}- A_2 - {\rm [Fe/H]}$ relations for fundamental-mode (RRab) and first-overtone mode (RRc) RR Lyrae stars (RRLs), respectively. The relations were calibrated based on pulsation periods and Fourier parameters of the RRL light curves in the {\it Gaia} $G$-band published in the {\it Gaia} Data Release 3 (DR3), and accurate spectroscopically measured metallicities available in the literature. We apply the feature selection algorithm to identify the most relevant parameters for the determination of metallicity. To fit the relations, we used the Bayesian approach, which allowed us to carefully take into account uncertainties in various parameters and the intrinsic scatter of the relations. The root mean squared errors of the predicted metallicity values in the training samples are 0.28~dex and 0.21~dex for RRab and RRc stars, respectively, comparable with the typical uncertainty of low/intermediate resolution spectroscopic metallicity measurements. 
We applied the new relations to measure individual metallicities and distances to $\sim$ 134,000 RRLs from the {\it Gaia} DR3 catalogue, as well as mean metallicities and distances to 38 Milky Way globular clusters. We also estimate the mean metallicity and distance to the Large Magellanic Cloud (LMC) and Small Magellanic Cloud (SMC): ${\rm [Fe/H]_{LMC} = -1.63\pm0.36}$ and $\mu_{\rm LMC}=18.55\pm0.18$~mag, ${\rm [Fe/H]_{SMC}=-1.86\pm0.36}$~dex and $\mu_{\rm SMC}=19.01\pm 0.17$~mag, respectively, in excellent agreement with previous measurements.

\end{abstract}




\section{Introduction}

RR Lyrae (RRLs) are pulsating variable stars that play a crucial role in stellar astrophysics. They are low-mass ($M<M_{\odot}$), old (age > 10 Gyr) stellar objects that populate the classical instability strip region of the horizontal branch (HB) in the colour–magnitude diagram (CMD). RRLs exhibit periodic variations in luminosity from about 0.2~mag up to more than a one magnitude in the visual band, occurring over a timescale ranging from a few hours to one day, which makes them easily detectable. They can be divided into three classes based on the pulsation modes: fundamental mode (RRab), first-overtone (RRc) and pulsating in both modes (RRd) stars. RRLs are abundant in the Milky Way (MW) halo and specific stellar systems, such as globular clusters (GCs), dwarf spheroidal (dSph) galaxies, and ultra-faint dwarf (UFD) galaxies, and are extensively used to study the properties of their host stellar systems (e.g. \citealt{Sesar2014}, \citealt{Molnar2015}, \citealt{Muraveva2020}, \citealt{Garofalo2021}). The presence of numerous RRLs in the MW halo provides an excellent opportunity to investigate the halo's overall shape, map the halo's substructures and constraint models of the formation and evolution of our Galaxy (e.g. \citealt{Drake2013}, \citealt{Belokurov2018}, \citealt{Iorio2019, Iorio2021}). 


RRLs are unique objects because their intrinsic properties, such as distance and metallicity ([Fe/H]), can be determined from easily observed photometric parameters (apparent magnitude, pulsation period, etc.). It is well known that one of the most direct and reliable methods to measure distance is trigonometric parallax, the determination of which requires accurate astrometric observations. The European Space Agency mission {\it Gaia} (\citealt{Prusti2016}) recently provided parallaxes for an unprecedented number of stars. However, the accuracy of the {\it Gaia} Early Data Release 3 (EDR3) parallaxes (0.02 - 0.03 mas for G $<$ 15 mag) drops dramatically for fainter objects, reaching values of 1.3~mas at G = 21 mag \citep{Brown2021}, which hampers a reliable estimation of distances directly from parallaxes for farther objects. Fortunately, RRLs can serve as reliable distance indicators helping to overcome {\it Gaia}’s limits. In fact, the distance to RRLs can be determined using the absolute magnitude-metallicity relation in the visual band ($M_V-{\rm [Fe/H]}$, e.g. \citealt{Clementini2003}, \citealt{Bono2003}), near-/mid-infrared period-luminosity-metallicity ($PLZ$) relations (e.g. \citealt{Longmore1986}, \citealt{Sollima2008}, \citealt{Madore2013}, \citealt{Muraveva2018a, Muraveva2018b}) and fundamental relations calibrated directly in the {\it Gaia} $G$, $G_{BP}$ and $G_{RP}$ bands (e.g. \citealt{Muraveva2018b}, \citealt{Garofalo2022}, \citealt{Li2023}). 

 At the same time, RRLs serve as useful metallicity tracers.
 The most direct method of measuring RRL metal abundances is from high-resolution (HR, R>=20,000) spectra, which provide metallicities with an accuracy of $\sim 0.1$ dex but require a large amount of telescope time. Nowadays, the metallicities measured from HR spectra exist for only a few hundred RRLs  (e.g. \citealt{Clementini1995}, \citealt{Lambert1996}, \citealt{For2011}, \citealt{Nemec2013}, \citealt{Pancino2015}, \citealt{Chadid2017}), even though this number increased significantly in the last few years (\citealt{Crestani2021}, \citealt{Gilligan2021}). The metallicity of RRLs can also be determined from low-resolution (LR) spectra through the $\Delta S$ method \citep{Preston1959}, which relies on the ratio between the equivalent width of Ca and H lines. The application of this method increases the number of RRLs with available metallicities to thousands (e.g. \citealt{Liu2020}, \citealt{Crestani2021}), but it is less accurate (typical error of $\sim$0.2 - 0.3 dex) and requires intermediate calibrations, which can introduce additional systematic errors. 

Fortunately, the metallicities of RRLs can also be determined using solely photometric observations. \cite{Jurcsik1996} found a linear relation between the metallicity of RRab stars and the Fourier parameter $\phi_{31}$ of their light curves in the $V$ band, along with the pulsation period. Later \cite{Morgan2007} produced a similar relation for RRc stars. Several authors calibrated these relations in other passbands (e.g. \citealt{Smolec2005}, \citealt{Nemec2013}, \citealt{Mullen2021}).
\cite{Hajdu2018} adopted a different approach to the problem of metallicity prediction, applying Machine Learning (ML) algorithms to the $K_s$-band light curves. Recently \cite{Dekany2021} used ML methods to derive the relation between Fourier parameters of the RRL light curves in the $I$-band, period and metallicity based on a sample of stars with metallicity known from HR spectroscopy.

Nowadays, it is a unique moment for RRLs thanks to data from large surveys, such as {\it Gaia} \citep{Prusti2016}. {\it Gaia}'s latest Data Release 3 (DR3) has provided the largest, most homogenous and parameter-rich catalogue of RRLs ever published (\citealt{Clementini2023}). Thus, calibration of the RRLs photometric metallicities directly from the $G$-band light curves becomes timely, especially in light of the coming {\it Gaia} DR4. To the best of our knowledge, direct relations between parameters of the RRL $G$-band light curves  and metallicity were provided so far by \cite{Iorio2021}, who used {\it Gaia} DR2 data and LR-spectroscopic metallicities from \cite{Layden1994}, and by \cite{Li2023}, who used {\it Gaia} DR3 light curves and a sample of RRLs with LR spectroscopy measurements from \cite{Liu2020}. At the same time, \cite{Dekany2022} used recurrent neural networks to predict the metallicity of RRLs from their $K_{s}$ and {\it Gaia} DR2 $G$-band light curves. 

In this paper, we derive new relations between periods, the Fourier parameters of the $G$-band light curve of RRLs and metallicity based on data from {\it Gaia} DR3 and HR/LR spectroscopic metal abundances available in the literature. To produce the new relations, we perform an accurate feature selection procedure and apply the Bayesian approach, allowing us to take into account possible systematics and properly estimate errors. We then apply our new relations to measure the photometric metallicities of 134,769 RRLs from the {\it Gaia} DR3 catalogue \citep{Clementini2023}. These metallicity estimates were used to calculate distances to RRLs applying the $G$-band luminosity-metallicity ($M_G - {\rm [Fe/H]}$) relation from \citet{Garofalo2022}. We analyse the MW structures traced by RRLs and provide new estimates of metallicity and distances to a number of GCs, hosting RRLs, and to the Magellanic Clouds. 
The paper is structured as follows: Section~\ref{sec:data} describes the dataset we used in our study. Section~\ref{sec:methods} outlines the applied methods. In Section~\ref{sec:met_val}, we perform different tests to evaluate the quality of the metallicities obtained in this study. In Section~\ref{sec:dist}, we estimate the distances to RRLs and analyse the MW structures as traced by RRLs. In Section~\ref{sec:mc} we study RRLs located in the Magellanic Clouds. Finally, Section~\ref{sec:summ} summarises our main results.

\section{Data}
\label{sec:data} 

\subsection{{\it Gaia} DR3 sample of RRLs}\label{subsec:gaia_sample}

\begin{figure*}
\includegraphics[width=8.5cm]{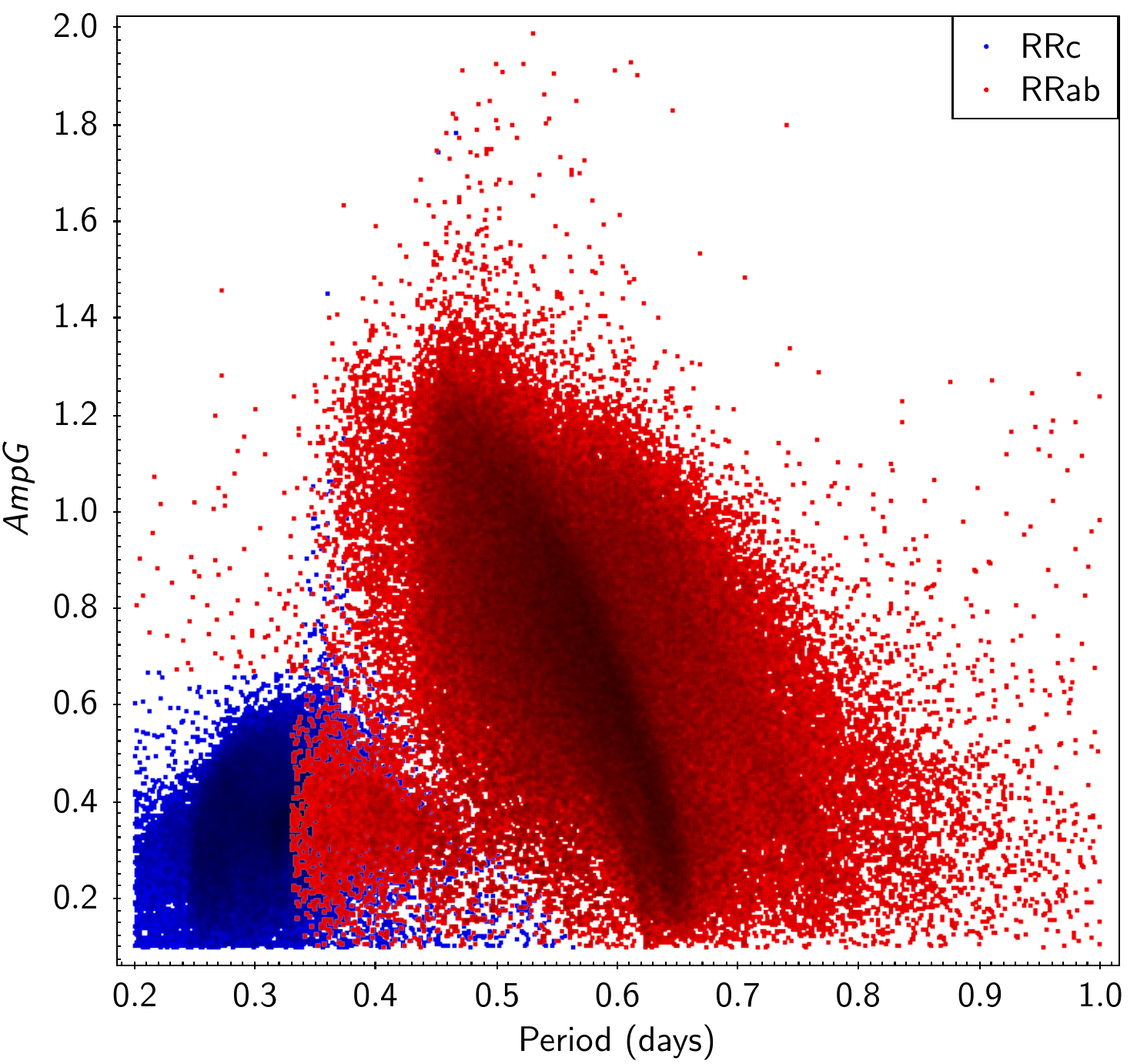}
\includegraphics[width=8.5cm]{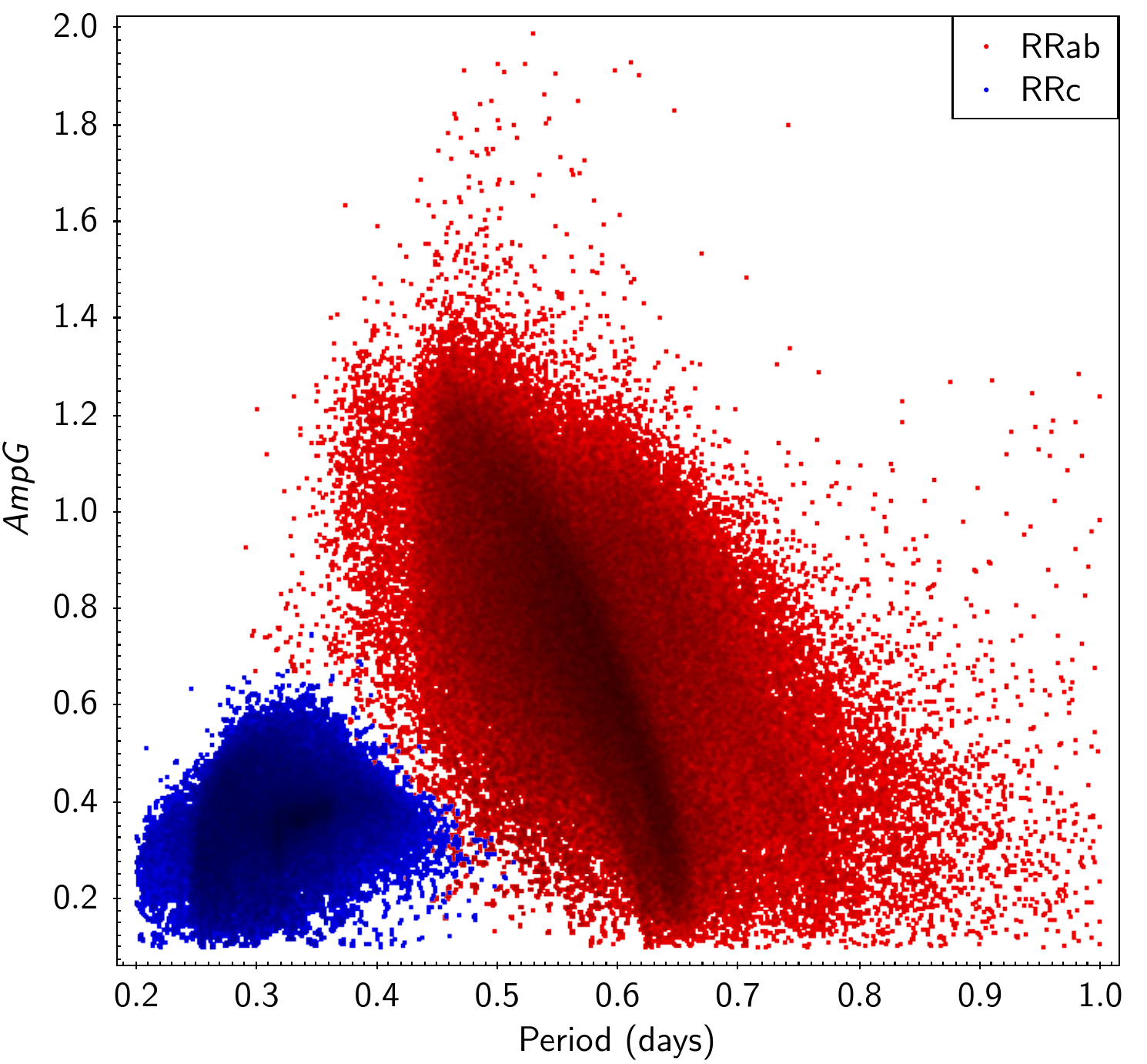}
    \caption{{\it Left panel:} Bailey diagram for RRab (red dots) and RRc (blue dots) stars in the {\it Gaia} DR3 catalogue. {\it Right panel:} Clean sample selected based on the RRLs' location on the Bailey diagram (GAIA-CAT-RRLS). See text for details.}
    \label{fig:bailey_gaia}
\end{figure*}

{\it Gaia} DR3 provides a catalogue of 271,779 RRLs observed during the initial 34 months of science operations and processed through the Specific Object Study pipeline for Cepheids and RR Lyrae stars (SOS Cep\&RRL, \citealt{Clementini2023}). This catalogue includes, among others, information on pulsation periods, peak-to-peak amplitudes of the $G$, $G_{BP}$ and $G_{RP}$ light curves, classification in pulsation mode, mean magnitudes computed as an intensity-average over the complete pulsation cycle, and parameters of the Fourier decomposition of the $G$-band light curves. During the final validation of the dataset \citet{Clementini2023} discarded  888 stars, mainly eclipsing binaries and objects with an uncertain classification. Thus, the final {\it Gaia} DR3 sample contains 270,891 RRLs (174,941 RRab, 93,944 RRc and 2006 RRd stars). We exclude RRd stars from our analysis and then cross-matched the remaining sample against the OGLE IV catalogue \citep{OGLE1, OGLE2}. There are 100,558 stars in common, for which we compared periods and classifications provided in both surveys. We found that classification in pulsation modes differ for 2505 stars, while there is a discrepancy in periods of more than 0.001 days for other 1520 sources. Since the sampling and temporal baseline of the OGLE light curves are usually better than those of {\it Gaia}, we decided to discard from our analysis the sources with classifications or periods different from the OGLE catalogue. 

The left panel of Figure~\ref{fig:bailey_gaia} shows the Bailey (amplitude in the $G$-band versus  pulsation period) diagram for RRab (red dots) and RRc (blue dots) stars in the remaining sample of 264,860 RRLs. As can be seen, some stars classified as RRab are situated within the zone typically occupied by RRc stars. These could be RRc stars misclassified as RRab stars or RRd stars with less than 40 transits observed by {\it Gaia}, for which the SOS pipeline does not search for the second periodicity \citep{Clementini2023}. Additionally, some RRab and RRc stars are positioned outside the expected loci, potentially indicating inaccuracies in their classification or periods provided in the {\it Gaia} DR3 catalogue. We cleaned the sample by selecting only RRLs located in the zones occupied by {\it Gaia} DR3 RRLs confirmed by OGLE. In order to do this we made a 2D Cartesian cross-match between period and the $G$-band amplitude of 264,860 RRLs shown in the left panel of Fig.~\ref{fig:bailey_gaia} and those parameters for the sample of {\it Gaia} DR3 RRLs confirmed by OGLE.  The right panel of Fig.~\ref{fig:bailey_gaia} shows the distribution of the clean sample of 258,696 RRLs (hereafter, GAIA-CAT-RRLS, Table~\ref{tab:cat}) on the Bailey diagram. We used this catalogue as a reference in the following analysis. 

\begin{table*}
	\centering
	\caption{Catalogues of RRLs used in the analysis}
	\label{tab:cat}
	\begin{tabular}{lcccccccc} 
		\hline
		Name & Source & RRab & RRc  & Total & [Fe/H] range  & [Fe/H] range  & Period range  & Period range  \\
         & & & & & RRab (dex) & RRc (dex) & RRab (days) & RRc (days) \\
		\hline
		HR-CAT-RRLS & \citet{Crestani2021}, & 126 & 24 & 150 & [-3.06, 0.11]  & [-2.60, -0.49] & [0.36, 0.96] & [0.22, 0.41] \\
  	 & \citet{Gilligan2021} & &  & &   &  & & \\
		LR-CAT-RRLS & \citet{Liu2020} & 797 & 452 & 1249 & [-2.92, -0.10] & [-2.87, -0.32] & [0.36, 0.93] & [0.22, 0.45] \\
		GAIA-CAT-RRLS & \citet{Clementini2023} & 169,024 & 89,672 & 258,696 & - & - & [0.29, 1.00] & [0.20, 0.51]\\
		\hline
	\end{tabular}
\end{table*}


\subsection{HR spectroscopic dataset}\label{subsec:HR}

\begin{figure}
\includegraphics[trim = 0 30 30 25, width=\columnwidth]{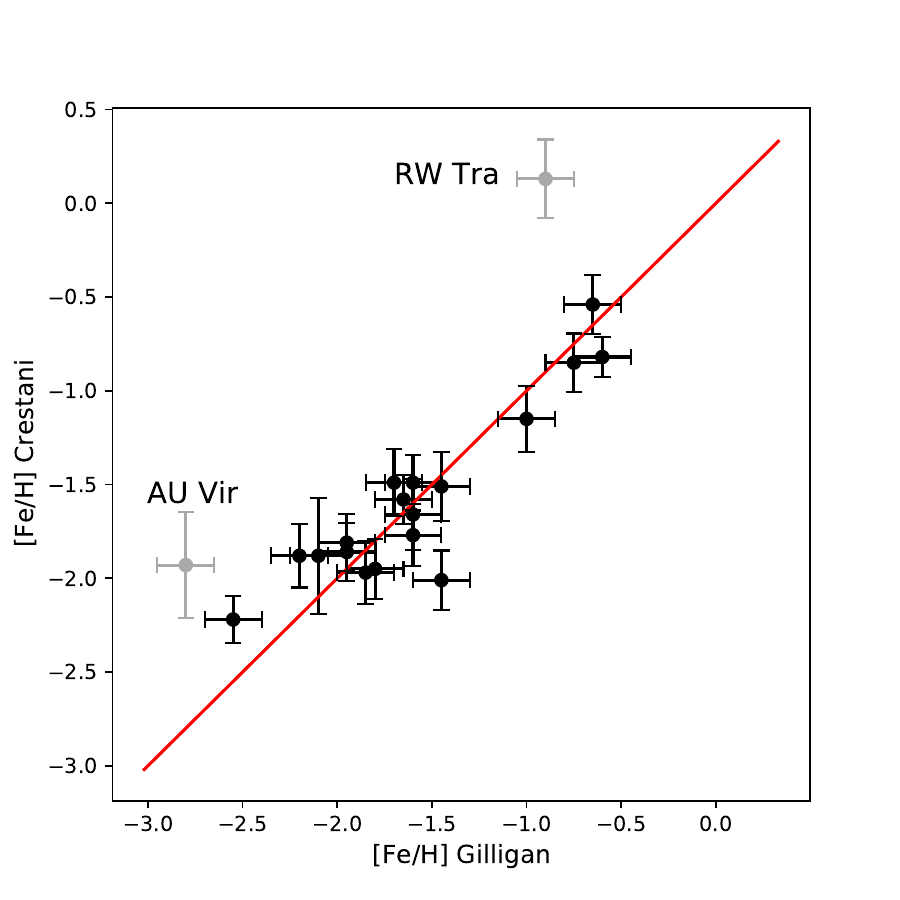}
\caption{Comparison between metallicities provided by \citet{Crestani2021} and \citet{Gilligan2021} for 20 RRLs in common}\label{fig:crestani_gilligan_comp}
\end{figure}

\begin{figure}

	\includegraphics[width=\columnwidth, trim=60 220 40 120]{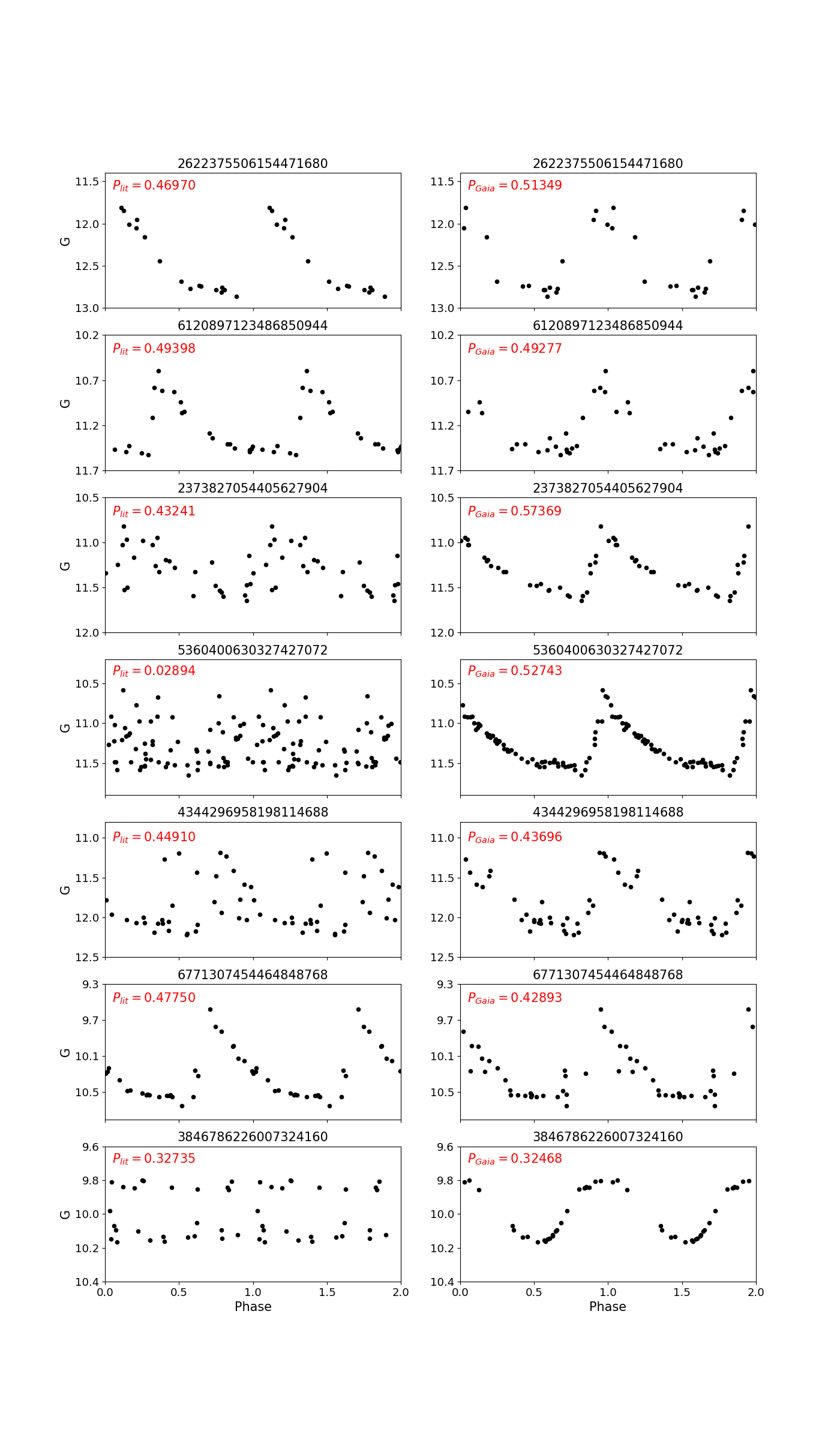}
    \caption{Light curves of sources, for which periods provided in the literature (\citealt{Crestani2021}, \citealt{Gilligan2021}) are in disagreement with those provided in {\it Gaia} DR3 catalogue. {\it Left panels:} Light curves phased using periods from the literature. {\it Right panel:} Light curves phased with the periods from the {\it Gaia} DR3 catalogue.}
    \label{fig:problem_lc}
\end{figure}

\begin{figure*}
\includegraphics[trim = 150 0 120 0,width = 17cm]{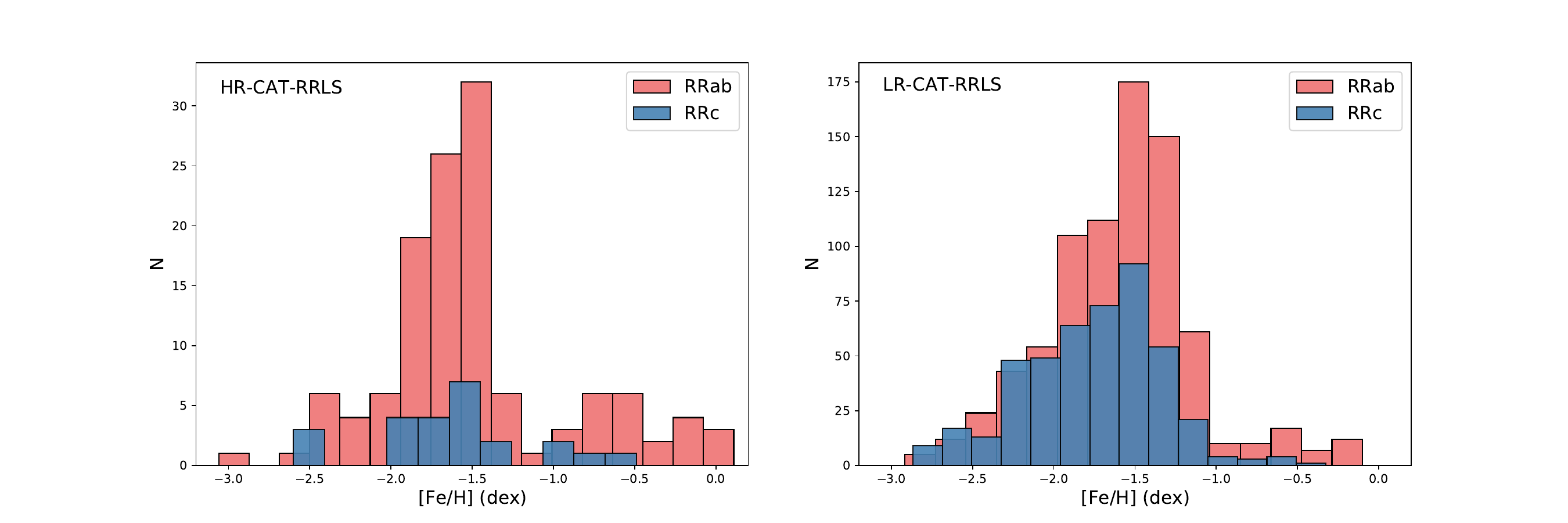}
    \caption{Metallicity distribution of RRLs in the HR-CAT-RRLS ({\it left panel}) and LR-CAT-RRLS ({\it right panel}) samples.}\label{fig:hist_met}
\end{figure*}

 \begin{figure*}
\includegraphics[trim =0 0 0 0, width=8 cm]{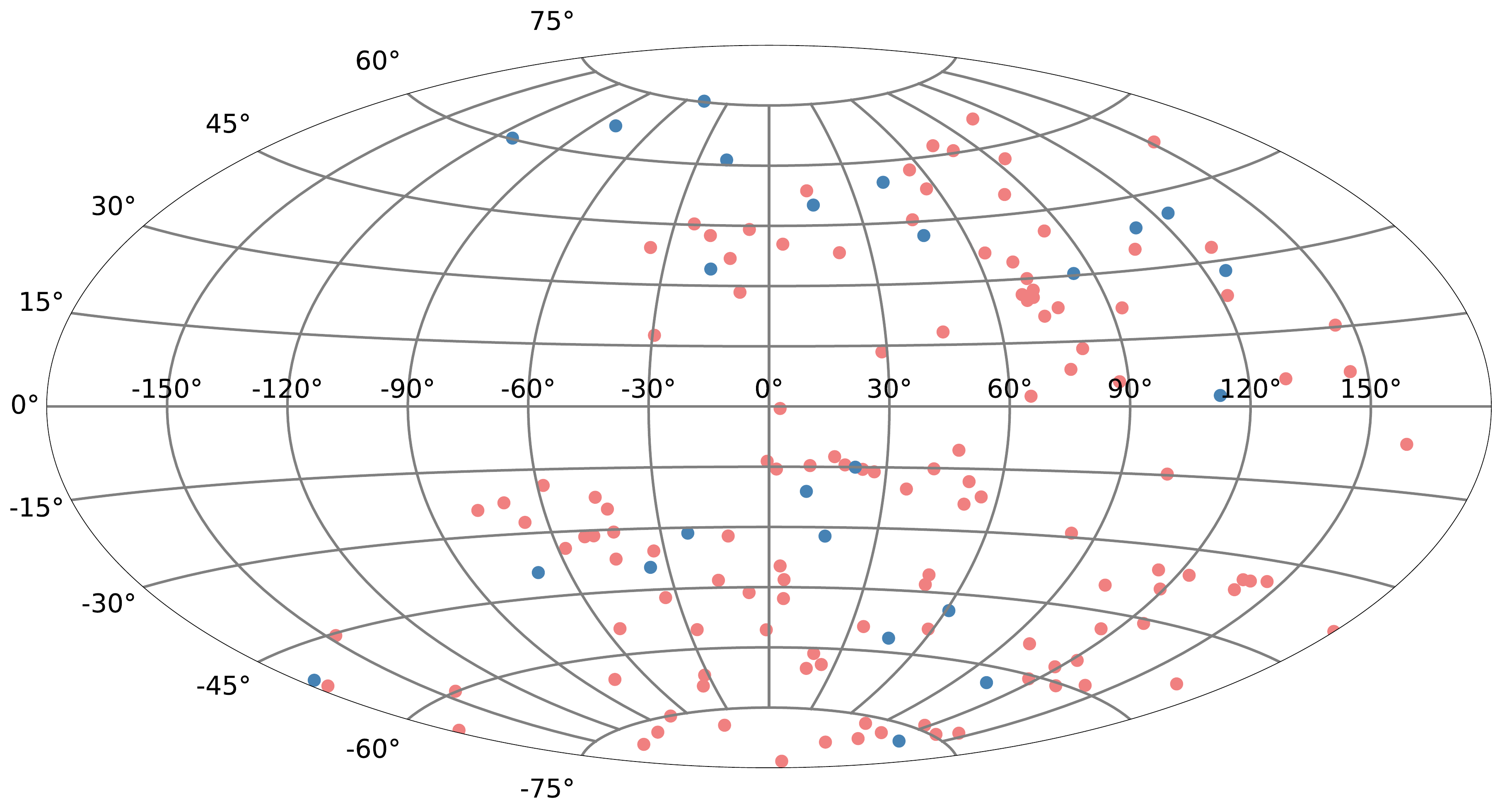}
\includegraphics[trim = 0 0 0 0, width=8 cm]
{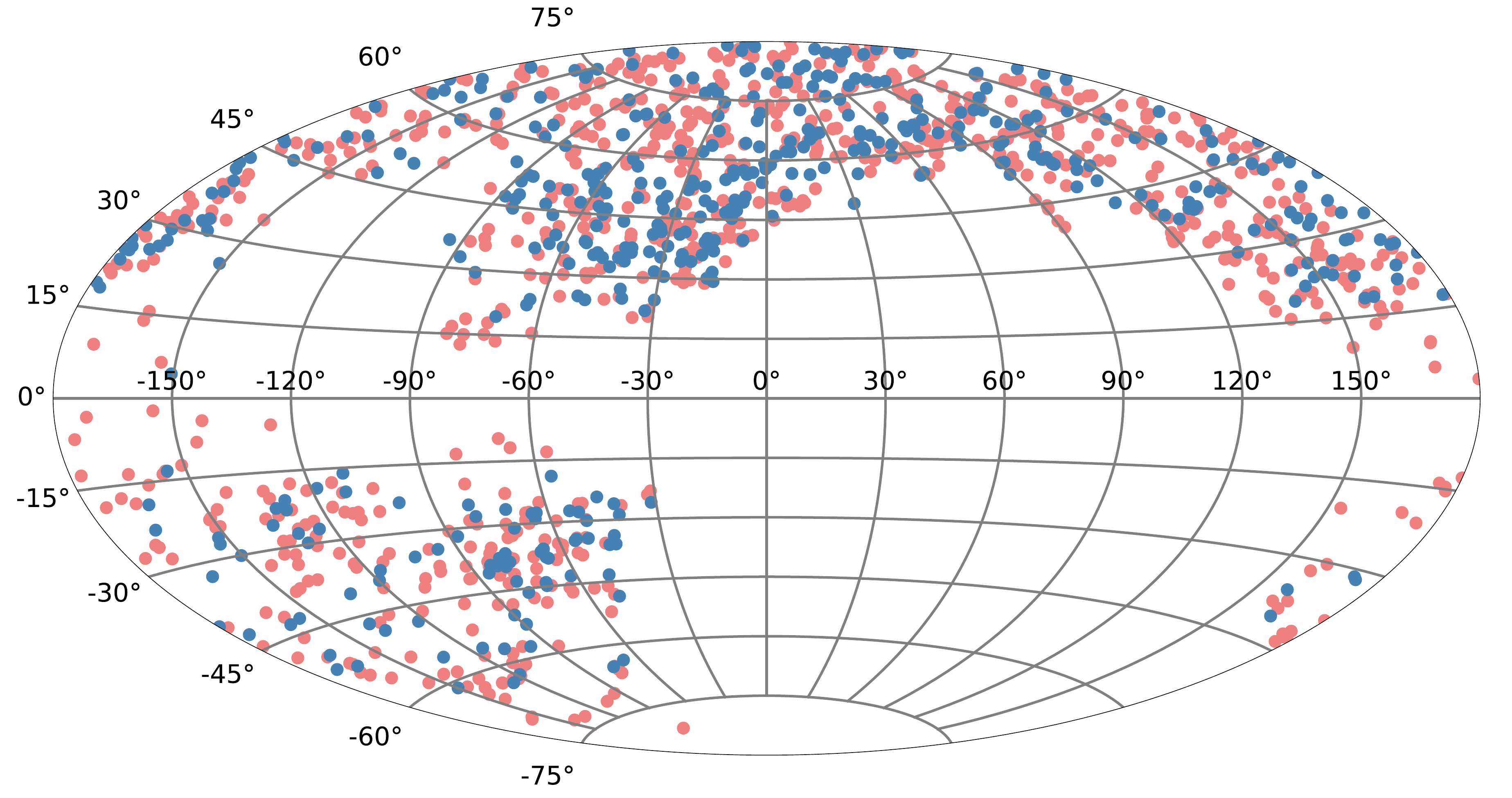}
    \caption{Spatial distribution of RRLs in the HR-CAT-RRLS ({\it left panel}) and LR-CAT-RRLS ({\it right panel}) samples.}\label{fig:sky_map}
\end{figure*}

To the best of our knowledge, the most extensive single catalogue of RRLs, for which metallicity is determined from HR spectroscopy, is published by \cite{Crestani2021}. Most RRL spectra were collected with the echelle spectrograph at du Pont (Las Campanas Observatory). This dataset was complemented with the HR metallicities estimated by \cite{For2011}, \cite{Chadid2017} and \cite{Sneden2017}. According to \cite{Crestani2021}, their measurements of metallicities, along with estimates from the three mentioned studies, can be treated as a single homogeneous sample, covering the range from $-3.0$ to 0.2~dex. The final calibrating sample of \cite{Crestani2021} includes 143 RRLs (111 RRab and 32 RRc stars). We used this catalogue as a reference sample for our training set. 

\cite{Crestani2021} calculated metallicities based on direct equivalent width measurements of both neutral (FeI) and ionized (FeII) iron lines in the HR spectra. When more than one measurement was available for a given star, the final metallicity value was calculated as the mean of all measurements, while the uncertainty was computed as the standard deviation of these measurements. However, this approach does not take into account individual uncertainties provided for each measurement. As a result, uncertainties in metallicities provided for some stars (Table~5 in \citealt{Crestani2021}) reach zero values. It is important to stress that robust uncertainty estimations are crucial for creating an accurate predictive model based on Bayesian analysis \citep{Dekany2021}. Thus, in our analysis, we recalculate the uncertainties of the metallicities provided by \cite{Crestani2021} by calculating the mean value of uncertainty of all available individual metallicity measurements (columns~8 and 10 in their Table~2) and adding it in quadrature to the uncertainty provided by \cite{Crestani2021} (column~4 of their Table~5). For the metallicities calculated by \cite{For2011}, \cite{Chadid2017} and \cite{Sneden2017}, we adopt 0.15 dex as a mean uncertainty of the individual metallicity measurements. Uncertainties calculated in this way span the range from 0.07 to 0.33~dex with a mean value of 0.17~dex.  

Recently, \cite{Gilligan2021} provided metallicities for 49 RRLs measured from HR spectra taken with the Southern African Large Telescope (SALT). The typical uncertainty of their metallicity measurements is 0.15~dex. In order to check if the metallicities determined by \citet{Gilligan2021} and \citet{Crestani2021} are on the same metallicity scale, we compare metal abundances calculated in both studies for 20 RRLs in common (Fig.~\ref{fig:crestani_gilligan_comp}). Apart from two obvious outliers (AU Vir and RW Tra), shown in grey, there is no visible offset between the two sets of metallicity measurements. The reason of discrepancy between the metallicity measurements for the two outliers is not clear. However, comparison with LR-spectroscopic metallicity estimates from \citet{Zinn2020} confirms the metallicity values provided by \citet{Crestani2021} for these two stars.  The weighted mean difference between \citet{Crestani2021} and \citet {Gilligan2021} metallicity measurements for the 20 RRLs in common is 0.05~dex, which is significantly smaller than the individual uncertainties. The difference has a negligible value of -0.005~dex if the two outliers shown in grey are excluded. We can conclude that the HR-spectroscopic metallicity measurements of \citet{Crestani2021} and \citet{Gilligan2021} can be considered to be on the same metallicity scale. Thus, we added 29 additional RRLs from \cite{Gilligan2021} to our sample of RRLs with HR spectroscopy metallicity estimates without applying any additional shift. Our sample of RRL calibrators with metal abundance from HR spectroscopy, thus, comprises 172 stars. 

We cross-matched our catalogue of RRLs with HR spectroscopy metal abundance against the GAIA-CAT-RRLS sample (Section~\ref{subsec:gaia_sample}) and found 162 RRLs in common. We compared the periods provided for these stars in the {\it Gaia} DR3 catalogue and in the literature (\citealt{Crestani2021}, \citealt{Gilligan2021}) and found that they differ by more than 0.001 days for seven sources. The light curves of these seven RRLs are shown in Fig.~\ref{fig:problem_lc}, where 
the periods from the literature were used to fold the data in the left panels and the {\it Gaia} DR3 periods in the right panels. 
Based on the visual inspection of the light curves, we confirm the period provided in the {\it Gaia} DR3 catalogue for the four sources with {\it Gaia} source\_id: 2373827054405627904, 4344296958198114688, 5360400630327427072 and 3846786226007324160. For the remaining three stars (2622375506154471680, 6120897123486850944, 6771307454464848768), the periods from the literature are correct (for the latter see also Table A.3 footnote (f) in \citealt{Clementini2023}). We exclude the latter three sources from our sample since their periods and, consequently, Fourier parameters provided in the {\it Gaia} catalogue could be incorrect.

We visually inspected the $G$-band light curves of all RRLs in the HR-spectroscopy sample. We found that nine stars have noisy light curves or gaps close to their light curves' maximum or minimum. Both issues could have affected the accurate estimation of the Fourier parameters of these RRLs; hence, we excluded them from our sample. 
Thus, our final sample of RRLs with metallicities determined from HR spectroscopy (hereafter, HR-CAT-RRLS) comprises 150 RRLs (Table~\ref{tab:cat_hr}). The metallicity distribution of the RRLs in the HR-CAT-RRLS is shown in the left panel of Fig.~\ref{fig:hist_met}, while the left panel of Fig.~\ref{fig:sky_map} shows their spatial distribution. The HR-CAT-RRLS includes RRLs from the bulge, disk, and halo of the MW. Table~\ref{tab:cat} contains more details on this sample.

\subsection{LR spectroscopic dataset}\label{subsec:lr}

\citet{Liu2020} published a catalogue of 5290 RRLs with metal abundances from low-resolution spectra of the LAMOST Experiment for Galactic Understanding and Exploration (LEGUE) and the Sloan Extension for Galactic Understanding and Exploration (SEGUE) surveys, estimated using the $\Delta S$ index \citep{Preston1959}. For this study, we selected only sources with spectral signal-to-noise ratio (S/N) larger than 50 from the \citet{Liu2020} sample. We then cross-matched this catalogue against the GAIA-CAT-RRLS sample (see Section~\ref{subsec:gaia_sample}) and found 1249 stars in common. Hereafter, we call this sample of RRLs with the metallicity estimated from LR spectroscopy LR-CAT-RRLS. See Table~\ref{tab:cat} for more details on this catalogue.

 The metallicity distribution of the RRLs stars in the LR-CAT-RRLS sample is shown in the right panel of Fig.~\ref{fig:hist_met}, while the right panel of Fig.~\ref{fig:sky_map} shows their spatial distribution.
Fig.~\ref{fig:sky_map} shows that RRLs in the HR-CAT-RRLS and LR-CAT-RRLS catalogues are distributed in a somehow complementary way. RRab stars in the HR-CAT-RRLS (Table~\ref{tab:cat}) span a slightly wider metallicity range than RRab stars in the LR-CAT-RRLS sample (Table~\ref{tab:cat}). Metallicity values measured using HR spectroscopy are also more accurate and do not imply intermediate calibrations, which can be an additional source of systematic errors. On the other hand, an important advantage of the LR-CAT-RRLS sample is that it includes many more RRab and, especially, RRc stars compared with the HR-CAT-RRLS sample.
 In the following analysis, we used the LR-CAT-RRLS sample when statistics over a large sample of stars was needed. The less numerous HR-CAT-RRLS catalogue is used instead when an accurate estimation of metallicity and related uncertainties is crucial.

\section{Methods}\label{sec:methods}

\begin{figure*}

	\includegraphics[trim = 0 0 0 0, width = 8cm]{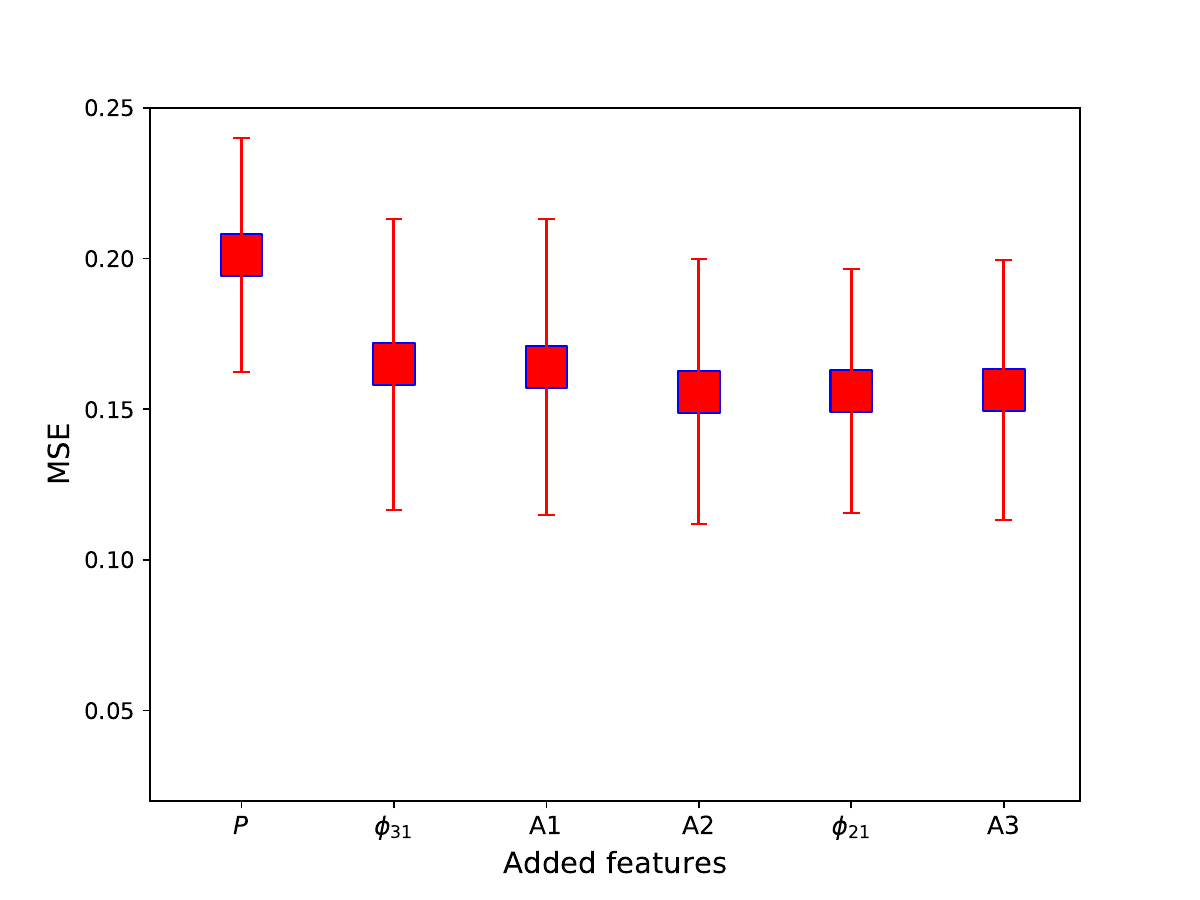}
    \includegraphics[trim = 0 0 0 0, width = 8cm]{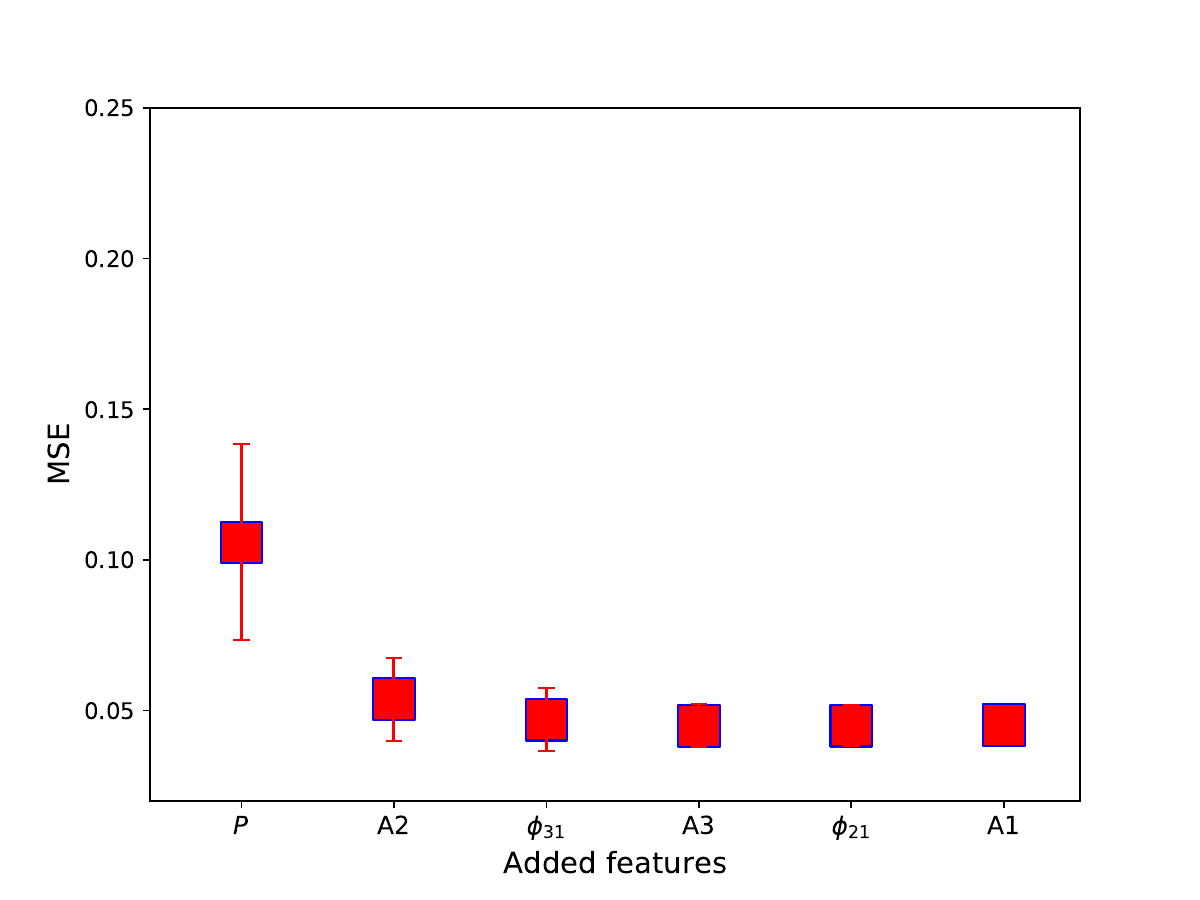}
    \caption{Distribution of the MSE estimated by cross-validation for RRab ({\it left panel}) and RRc ({\it right panel}) stars. The X axes show the features subsequently added to the subset.}
    \label{fig:ab_score}
\end{figure*}

The methodological approach used in this study is similar to the analysis performed by \cite{Dekany2021}, who applied the Sequential Feature Selection (SFS) and the Bayesian regression to find the photometric metallicities of RRLs from their $I$-band light curves. In our analysis, we divide the problem of predicting the metallicity of RRLs into two steps: (1) feature selection and (2) application of the Bayesian regression to derive a relation between period, Fourier parameters of the $G$-band light curves and metallicities.

\subsection{Feature selection}\label{subsec:sfs}

{\it Gaia} SOS Cep\&RRL pipeline \citep{Clementini2023} represents the $G$-band light curves of RRLs using a Fourier expansion in cosine functions:

\begin{equation}
mag(t_j) = zp + \sum[A_{i}\cos(i\times2\pi\nu_{max}t_{j}+\phi_{i})],
	\label{eq:quadratic}
\end{equation}

where $zp$ is the zero-point, $\nu_{max}$ is the pulsation frequency ($\nu_{max}$=1/P , where P is the pulsation period), $i$ is the number of harmonics used to model the $G$-band light curve, $A_i$ are amplitudes and $\phi_i$ are phases of the Fourier decomposition. The \texttt {vari\_rrlyrae} table in the {\it Gaia} DR3 catalogue provides amplitudes and phases of the Fourier decomposition of RRL $G$-band light curves, as well as phase parameters $\phi_{21} = \phi_2-2*\phi_1$ and $\phi_{31} = \phi_{3} - 3*\phi_{1}$, and their respective uncertainties \citep{Clementini2023}. While, historically photometric metallicities of RRLs were obtained from the pulsation period and $\phi_{31}$ parameter (e.g. \citealt{Jurcsik1996}, \citealt{Nemec2013}), in some following studies additional Fourier parameters, such as amplitudes $A_1$ and $A_2$ \citep{Dekany2021} or their ratio $R_{21} = A_2/A_1$ \citep{Li2023} were used. In our study, we apply the Sequential Feature Selection (SFS) algorithm implemented in the \texttt{scikit-learn} software library \citep{scikit-learn} to identify the most relevant features that, in the best way, predict the metallicity of  RRab and RRc stars. Since a large sample of RRLs is needed to identify the features correctly, we apply the SFS to the LR-CAT-RRLS sample (Section~\ref{subsec:lr}). We adopt the forward-SFS algorithm, which is a sequential process starting with an empty set of features and iteratively adding one feature at the time that provides the maximum improvement in model performance. The performance evaluation at each step is done through a cross-validation (CV) algorithm. Namely, after adding each feature, the training sample is divided into five subsets. The model (linear regression in our case) is trained on four subsets, while the fifth is used to evaluate the model performance by computing the mean squared error (MSE). This procedure is repeated for the remaining four subsets, and the final MSE is then the average of the values computed for each subset. Based on this performance evaluation, the algorithm decides which feature should be added to the subset of relevant features. The main advantage of the SFS algorithm is that it explores different feature combinations and evaluates their impact on the model performance. It allows us to identify the most informative features and discard irrelevant or redundant ones, potentially improving the model's efficiency.

We selected pulsation period P, $\phi_{21}$, $\phi_{31}$, $A_1$, $A_2$ and $A_3$ as the initial set of features and run the SFS algorithm to select the most relevant features for predicting the metallicity of 670 RRab stars from the LR-CAT-RRLS, for which all six features were available. The upper section of Table~\ref{tab:features} and left panel of Fig.~\ref{fig:ab_score} show the distribution of the MSE for different subsets of features. Based on our analysis, the most important features for metallicity prediction of RRab stars are $P$ and $\phi_{31}$, while adding $A_1$ does not improve the MSE performance beyond its uncertainty. Considering that adding more features could cause additional noise, we decide to limit the set of features for RRab stars to [$P$, $\phi_{31}$], which is in agreement with previous studies (e.g. \citealt{Jurcsik1996}, \citealt{Nemec2013}). 

 We repeated the same analysis for 226 RRc stars in the LR-CAT-RRLS, for which all six features were available. Results are shown in the bottom section of Table~\ref{tab:features} and the right panel of Fig.~\ref{fig:ab_score}. The most important features for estimating RRc metallicity are $P$ and $A_{2}$, while adding $\phi_{31}$ slightly improves the model performance. We, thus, select the set of features [$P$, $\phi_{31}$, $A_{2}$] for metallicity prediction of RRc stars.

\begin{table}
	\centering
	\caption{Performance of the linear model for different feature sets.}\label{tab:features}
	\label{tab:feature}
	\begin{tabular}{lcc} 
		\hline
  Features & MSE  & $\sigma(MSE)$ \\
  \hline
  \multicolumn{3}{c}{RRab} \\
  \hline
{P} & 0.201 & 0.039 \\
{P, $\phi_{31}$} & 0.165 & 0.048 \\
{P, $\phi_{31}$, $A_{1}$} & 0.164 & 0.049 \\
{P, $\phi_{31}$, $A_{1}$, $A_{2}$} & 0.156 & 0.044\\
{P, $\phi_{31}$, $A_{1}$, $A_{2}$, $\phi_{21}$} & 0.156 & 0.041 \\
{P, $\phi_{31}$, $A_{1}$, $A_{2}$, $\phi_{21}$, $A_{3}$} & 0.156 & 0.043\\
		\hline
\multicolumn{3}{c}{RRc} \\
\hline
{P} & 0.106 & 0.032 \\
{P, $A_{2}$} & 0.054 & 0.014 \\
{P, $A_{2}$, $\phi_{31}$} & 0.047 & 0.010 \\
{P, $A_{2}$, $\phi_{31}$, $A_{1}$} & 0.045 & 0.007 \\
{P, $A_{2}$, $\phi_{31}$, $A_{1}$, $A_{3}$} & 0.045 & 0.007 \\
{P, $A_{2}$, $\phi_{31}$, $A_{1}$, $A_{3}$, $\phi_{21}$} & 0.045 & 0.006 \\

\hline
	\end{tabular}
\end{table}

\subsection{Bayesian approach}\label{subsec:bayes}

\begin{figure*}
\includegraphics[trim = 0 0 0 0, width=15cm]{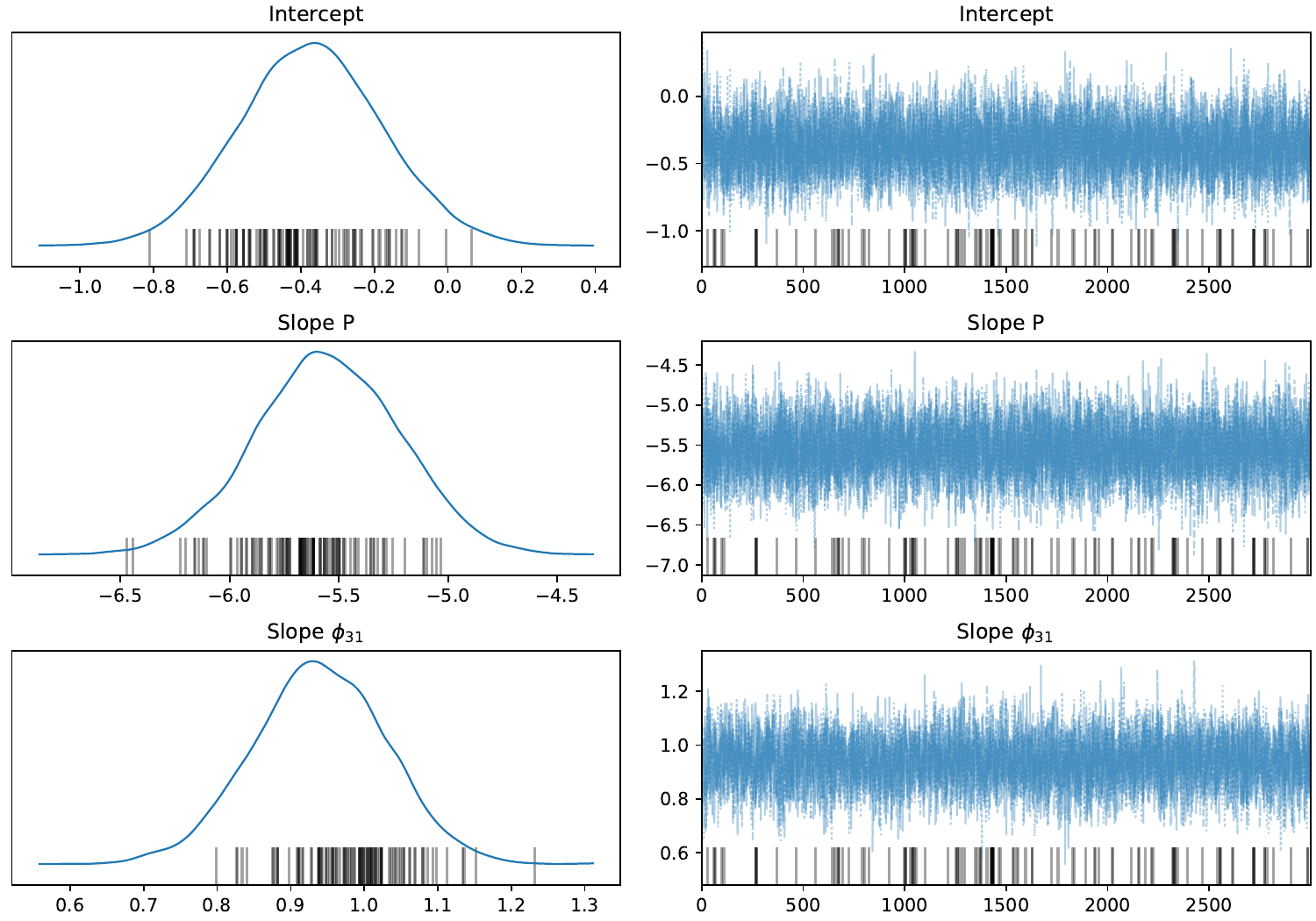}
\caption{{\it Left panel}: Posterior distributions of the parameters of the predictive metallicity model for RRab stars (Eq.~\ref{eq:rrab}). {\it Right panel}: Values of the parameters over the course of the MCMC sampling process. Each point on the X-axis corresponds to a specific sample from the posterior distribution}\label{fig:trace_ab} 
\end{figure*}

To predict the photometric metallicity of RRLs from the period and Fourier parameters of the $G$-band light curves, we use the Bayesian fitting approach described in detail by \cite{Dekany2021} and \cite{Muraveva2018a}. As discussed in Section~\ref{subsec:sfs}, the most important features to predict RRab metallicities are the period and the $\phi_{31}$ parameter. As a training set for the Bayesian predictive model, we used 121 RRab stars from the HR-CAT-RRLS, for which both features were available. We used the \texttt{pymc3} software library to fit the model. Posterior distributions were calculated applying Hamiltonian Markov chain Monte Carlo (MCMC) simulations using the No-U-Turn Sampler (NUTS, \citealt{Hoffman2011}). Fig.~\ref{fig:trace_ab} shows the posterior distribution of the parameters of the model and the values of the parameters over the course of the MCMC sampling process. The uncertainties in the period, $\phi_{31}$ and metallicity, as well as the intrinsic scatter of the relation, were taken into account. Metallicity uncertainties were estimated as described in Section~\ref{subsec:HR}. Uncertainties in periods provided in the {\it Gaia} DR3 catalogue \citep{Clementini2023} are negligibly small ($\sim10^{-6}$). Nevertheless, in Section~\ref{subsec:HR}, we found that {\it Gaia} periods agree with the literature values within 0.001 days. Thus, we adopt this value as the period's uncertainty. Uncertainty in the Fourier parameter $\phi_{31}$ was taken from {\it Gaia} DR3 catalogue (\citealt{Clementini2023}).

Performing the fitting procedure, we found two strong outliers: SS~Gru and WZ~Hya. SS~Gru is the longest-period star in our sample. In some studies (e.g. \citealt{Plachy2021}) it is classified as anomalous Cepheid. 
WZ~Hya has the largest values of the $\phi_{31}$ parameter and corresponding uncertainty $\phi_{31} = 3.41\pm0.81$ among all stars in the sample. This value of $\phi_{31}$ could probably be incorrect, especially considering a relatively small number of transits (N=27) used for fitting the light curve. We decide to exclude both stars from our sample as strong outliers. 

We found the following relation for the metallicity of RRab stars:
\begin{equation}
{\rm [Fe/H]} = (-5.55 \pm 0.33)P + (0.94 \pm 0.09)\phi_{31} -  (0.37 \pm 0.20)\label{eq:rrab}
\end{equation}
The intrinsic dispersion of the fit is $\sigma =0.21$ dex. The root mean squared error (RMSE) of the predicted metallicities in the training sample is 0.28~dex.

We performed the same analysis for RRc stars. As shown in Section~\ref{subsec:sfs}, the most important features to predict the metallicity of RRc stars are period, $A_2$ and $\phi_{31}$ Fourier parameters. The HR-CAT-RRLS sample contains only 14 RRc stars, for which all three parameters are available, which is insufficient to construct a reliable predictive model. Thus, we used RRc stars from the LR-CAT-RRLS sample to fit the relation. However, metallicity values of RRc stars in LR-CAT-RRLS from \citet{Liu2020} are not on the metallicity scale adopted by \citet{Crestani2021}, that we used to calibrate the relation of RRab stars from HR-CAT-RRLS (Eq.\ref{eq:rrab}). \citet{Crestani2021} in their analysis, performed a comparison between their metallicity estimates and values provided by \citet{Liu2020} for 2634 stars in common and found a mean difference of $\Delta = -0.21$~dex for RRc stars. To homogenise the metallicity scales adopted for RRab stars and RRc stars, we add a shift of $+0.21$~dex to the metallicity values of RRc stars in the LR-CAT-RRLS sample. In this way, the relations calculated for RRab and RRc stars in our study are on the same metallicity scale adopted by \citet{Crestani2021}. 

\begin{figure*}
\includegraphics[trim = 0 0 0 0, width=15cm]{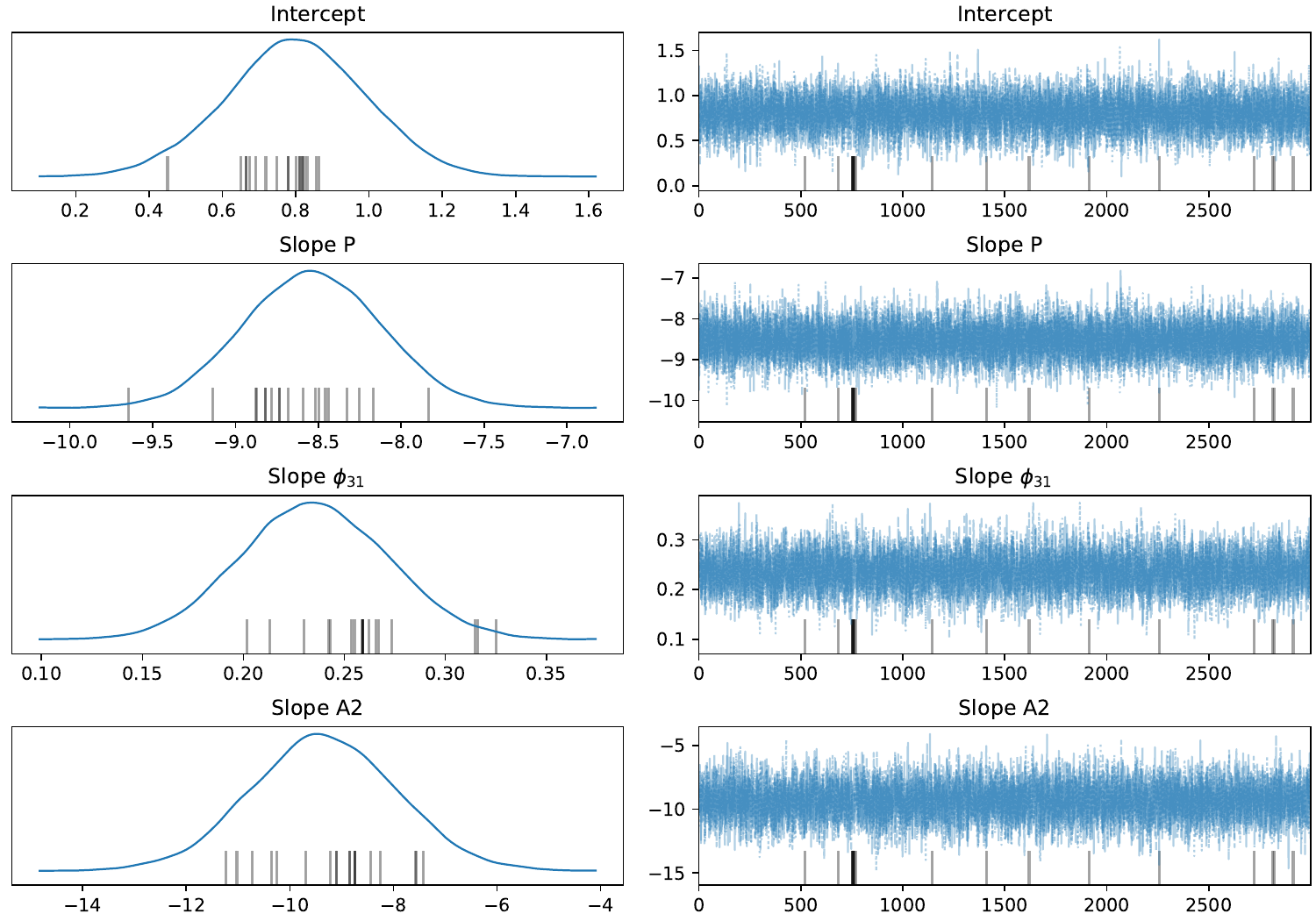}
\caption{{\it Left panel}: Posterior distributions of the parameters of the predictive metallicity model for RRc stars (Eq.~\ref{eq:rrc}). {\it Right panel}: Values of the parameters over the course of the MCMC sampling process. Each point on the X-axis corresponds to a specific sample from the posterior distribution}\label{fig:trace_c} 
\end{figure*}

All three features [P, $A2$, $\phi_{31}$] are available for 226 RRc stars from the LR-CAT-RRLS. We applied the Bayesian fitting method and found the following relation for the metallicity of RRc stars:

\begin{equation}
\begin{split}
{\rm [Fe/H]} = (-8.54 \pm 0.42)P + (0.23 \pm 0.04)\phi_{31} -  (9.34 \pm 1.41)A_2 +
\\ (0.80 \pm 0.19)\label{eq:rrc}
\end{split}
\end{equation}
The intrinsic scatter of the fit is $\sigma =0.19$~dex, while RMSE is 0.21~dex.  Fig.~\ref{fig:trace_c} shows the posterior distribution of the parameters of the model and the values of the parameters over the course of the MCMC sampling process.
In the next section, we perform a validation of our newly derived relations, while in Appendix~\ref{sec:xgboost} we use the XGBoost regressor to 
explore the potential existence of more complex relationships between light curve parameters and metallicity of RRLs.

\section{Metallicity validation}\label{sec:met_val}

\begin{figure}
\includegraphics[trim = 20 30 60 20,width = 8cm]{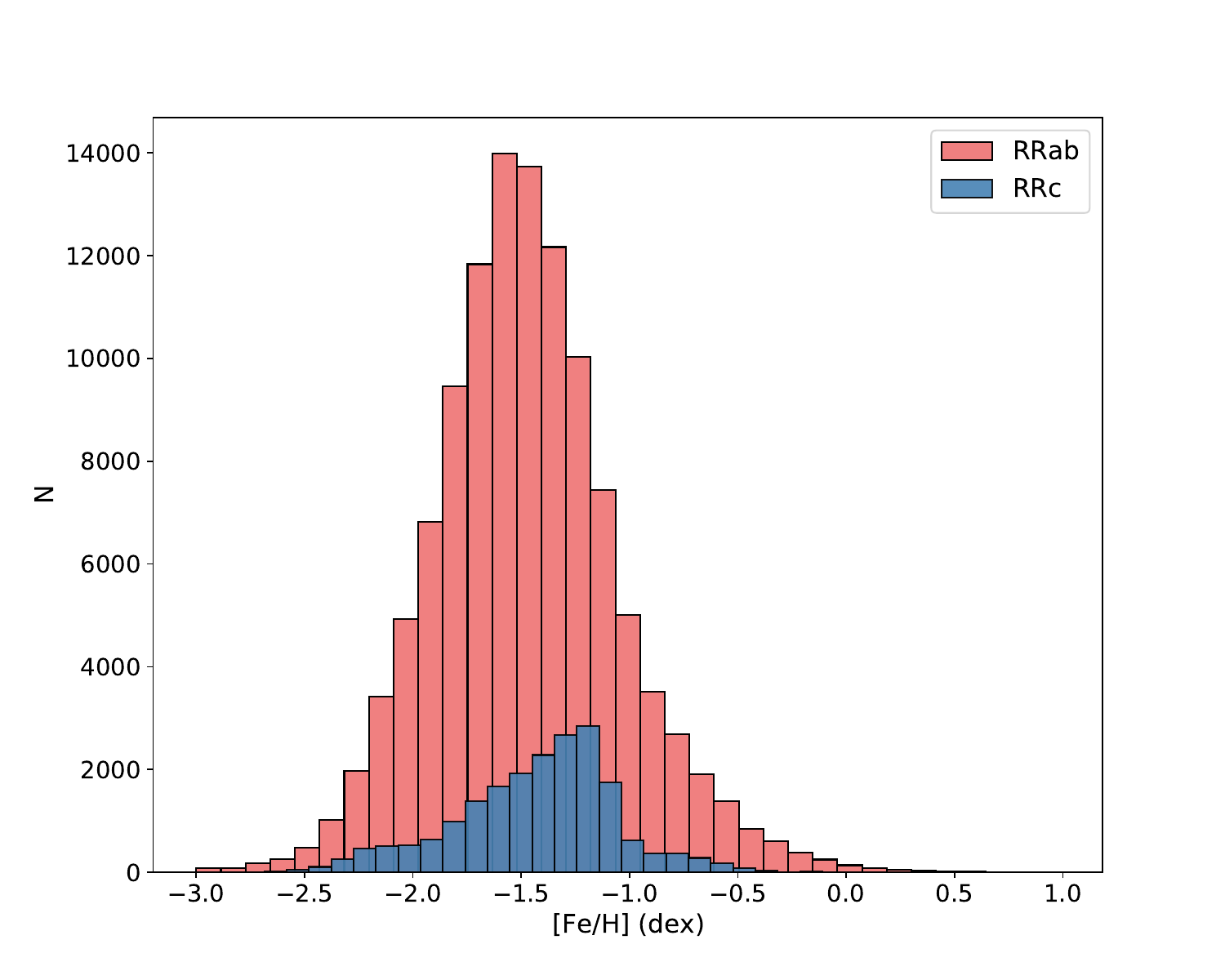}
    \caption{Distribution of metallicities of 134,769 RRLs estimated using Eqs.~\ref{eq:rrab}-\ref{eq:rrc}\label{fig:hist_met_our}}
\end{figure}

\begin{figure*}
\includegraphics[width = 17cm]{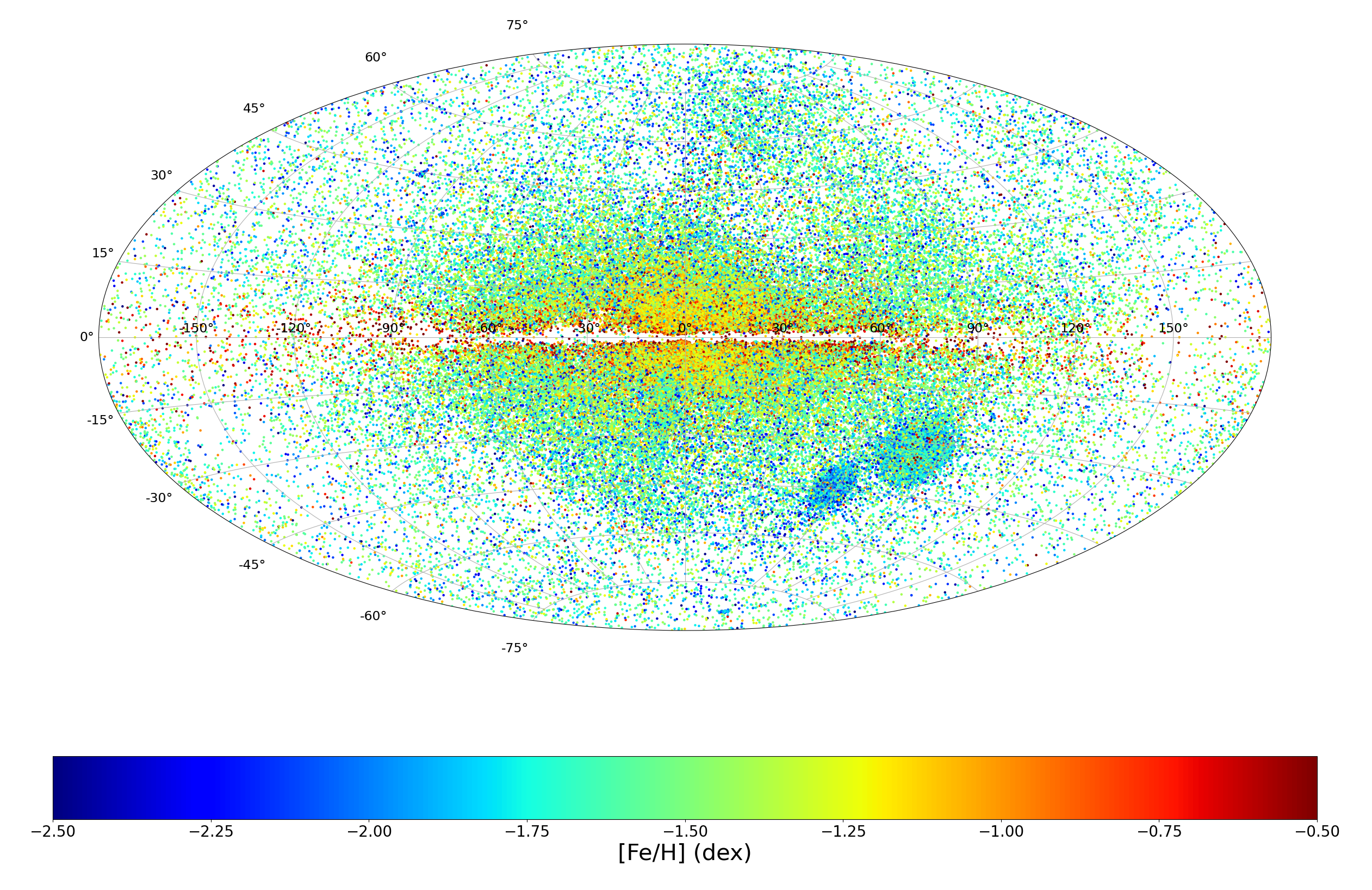}
    \caption{Sky distribution of 134,769 RRLs from the GAIA-CAT-RRLS sample. Sources are colour-coded according to the metal abundance obtained using Eqs.~\ref{eq:rrab} and \ref{eq:rrc}. There are 803 RRLs with $-$3.0<[Fe/H]<$-$2.5 dex (all shown in dark blue) and 2557 RRLs with [Fe/H]>$-$0.5 dex (shown in dark red).}
    \label{fig:sky_met}
\end{figure*}

\begin{figure}
\includegraphics[width = \columnwidth, trim = 30 50 50 40]{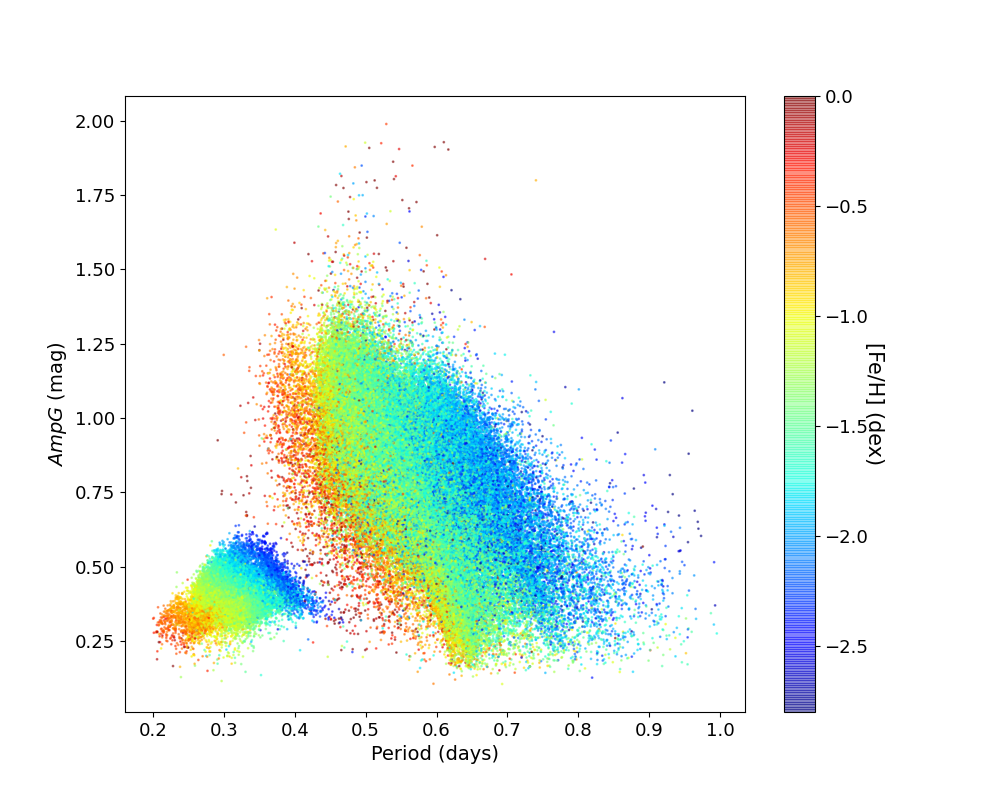}
    \caption{Amplitude in the $G$-band versus period distribution of the 134,769 RRLs from the GAIA-CAT-RRLS sample, colour-coded according to the photometric metallicity obtained using Eqs.~\ref{eq:rrab} and \ref{eq:rrc}.}
    \label{fig:bailey}
\end{figure}

\begin{figure*}
\includegraphics[width = 18cm, trim = 0 50 0 0]{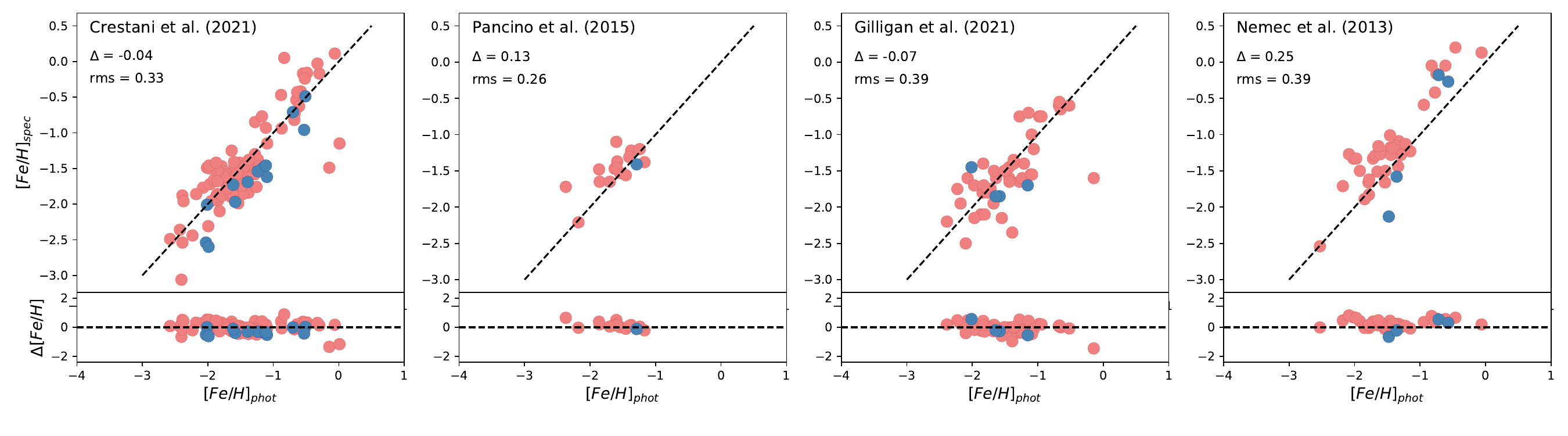}
    \caption{Comparison between photometric metallicity values calculated using Eqs.~\ref{eq:rrab}-\ref{eq:rrc} and HR-spectroscopic metallicities from the literature for RRab (red circles) and RRc (blue circles) stars. Black dashed lines in the upper panels represent the one-to-one relations. The bottom panels show residuals ${\rm \Delta [Fe/H]=[Fe/H]_{spec} - [Fe/H]_{phot}}$.}
    \label{fig:comp_lit_hr}
\end{figure*}

\begin{figure*}
\includegraphics[width = 18cm, trim = 0 50 0 0]{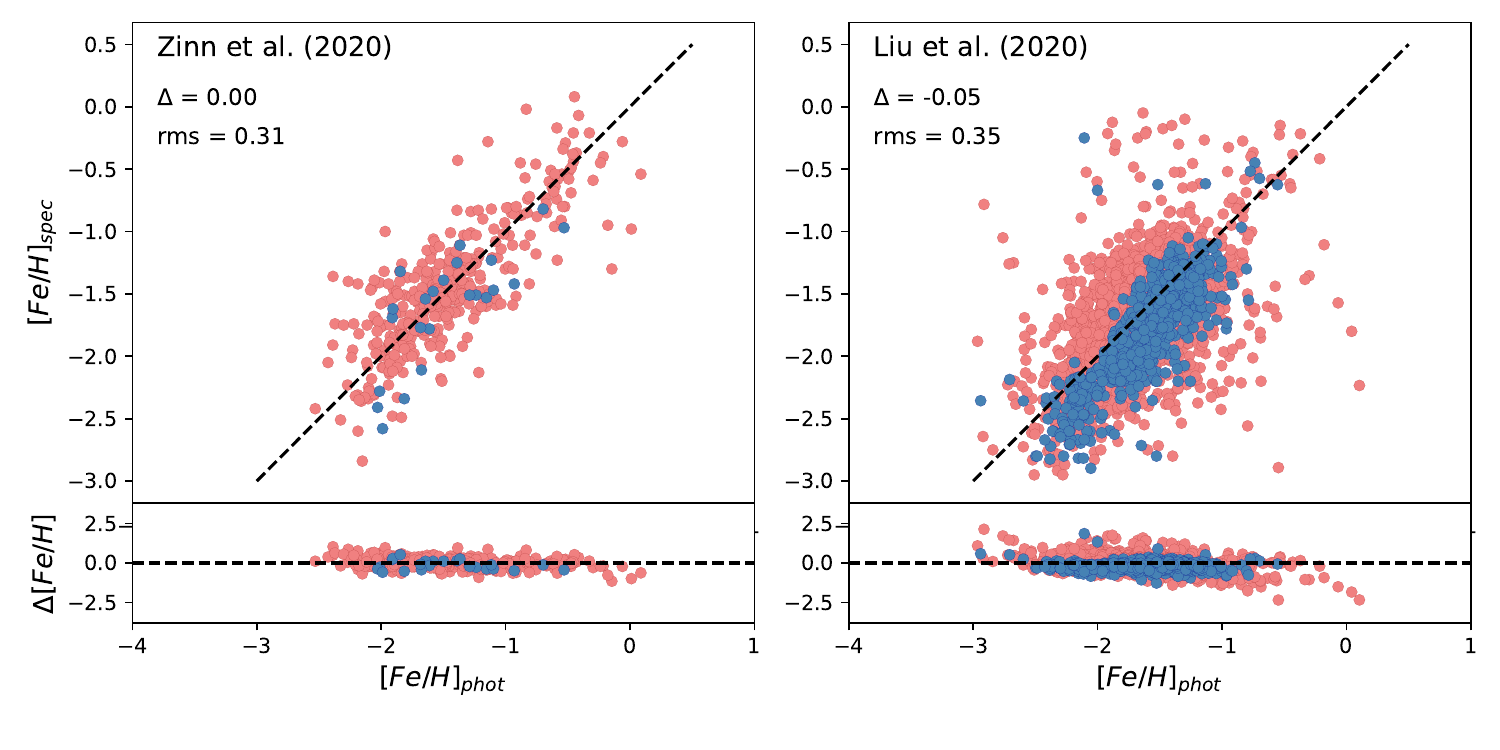}
    \caption{Comparison between photometric metallicity values calculated using Eqs.~\ref{eq:rrab}-\ref{eq:rrc} and LR-spectroscopic metallicities from different studies in the literature for RRab (red circles) and RRc (blue circles) stars. Black dashed lines in the upper panels represent the one-to-one relations. The bottom panels show the residuals ${\rm \Delta[Fe/H]=[Fe/H]_{spec} - [Fe/H]_{phot}}$.}
    \label{fig:comp_lit_lr}
\end{figure*}

\begin{figure*}
\includegraphics[width = 17cm]{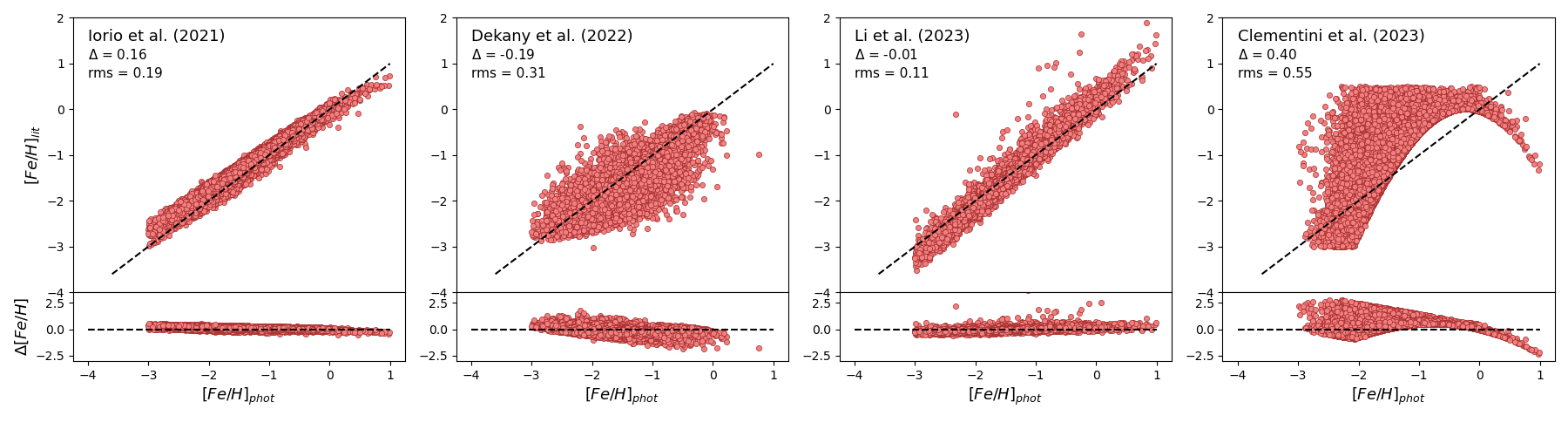}
\includegraphics[width = 17cm]{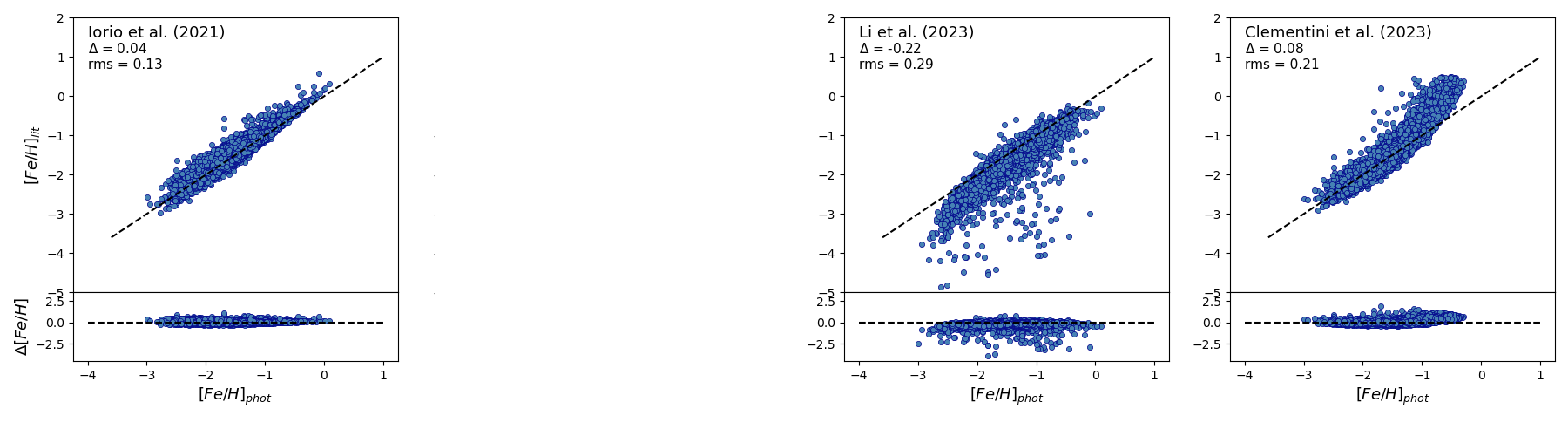}
    \caption{Comparison between photometric metallicity values calculated using Eqs.~\ref{eq:rrab}-\ref{eq:rrc} and metallicities from different studies in the literature calculated using {\it Gaia} data for RRab (red circles) and RRc (blue circles) stars. Black dashed lines in the upper panels represent the one-to-one relations. The bottom panels in each plot show the residuals ${\rm \Delta[Fe/H]=[Fe/H]_{lit} - [Fe/H]_{phot}}$.}
    \label{fig:comp_lit_gaia}
\end{figure*}

We have applied our newly derived relations (Eqs.~\ref{eq:rrab}-\ref{eq:rrc}) to compute the photometric metallicities of 135,033 RRLs in the GAIA-CAT-RRLS sample, for which all needed parameters were available.
We estimated the uncertainties in metallicity by employing a Monte Carlo simulation approach. For each star, 1000 iterations were performed. The random values were sampled from the error distributions of the period, Fourier parameters and coefficients of the relations (Eqs.~\ref{eq:rrab} - \ref{eq:rrc}), while also simulating the intrinsic dispersion in metallicity ($\sigma$). 
The collection of the metallicity values obtained from the Monte Carlo simulation allowed us to estimate the uncertainties in metallicity.

Applying linear relations to estimate metallicities, we found that for 264 stars (0.002\% of the sample), the metallicity reaches unreliable values from -3 up to -5.65~dex at the metal-poor end, and from 1 up to 2.86~dex at the metal-rich end of the metallicity distribution. This is mostly due to unreliable estimates of the period and/or Fourier parameters for a handful of stars. We exclude these stars and remain with a sample of 134,769 RRLs with estimated metallicity values (Table~\ref{tab:cat_gaia}). The metallicity distribution of these RRLs is shown in Fig.~\ref{fig:hist_met_our} and spans a total range of  -3.0<[Fe/H]<+1.0~dex. In particular, there are 178 RRLs with [Fe/H]>+0.1 dex in the sample, hence they are outside the metallicity range covered by our  RRL calibrators (-3.06 <[Fe/H]<+0.11 dex), Table~\ref{tab:cat}). Their metal abundance should be taken with care.
Fig~\ref{fig:sky_met} shows the sky distribution of 134,769 RRLs in our sample, while Fig.~\ref{fig:bailey} shows their Bailey diagram. In both plots the colour encodes the stars' metallicity.

\subsection{Comparison with the literature}\label{subsec:comp_lit}

\begin{table*}
	\caption{Parameters for 38 GCs containing more than 5 RRLs, for which we estimated photometric metallicity using Eqs.~\ref{eq:rrab}-\ref{eq:rrc}: (1) name of the GC; (2) and (3) mean value and standard deviation of the metallicity estimates of the RRLs in each GC on \citet{Crestani2021} metallicity scale calculated using Eqs.~\ref{eq:rrab}-\ref{eq:rrc}; (4) and (5) metallicity and corresponding uncertainty of GCs from \citet{Carretta2009}; (6) and (7) mean value and standard deviation of the GC's distance moduli, calculated using individual distance estimates to the RRLs in each GC; (8) reddening value from \citet{Harris2010}; (9) number of RRLs in the GC with photometric metallicity estimates.}
	\label{tab:gc}
	\begin{tabular}{lcccccccc} 
		\hline
		Name & ${\rm [Fe/H]_{RRLS}}$ & $\sigma {\rm [Fe/H]_{RRLS}}$& ${\rm [Fe/H]_{C09}}$& $\sigma {\rm [Fe/H]_{C09}}$& $\mu$ & $\sigma \mu$ & E(B-V) & $N_{\rm {RRLS}}$\\
        & (dex) & (dex) & (dex) & (dex) & (mag) & (mag) & (mag) & 
  \\
		\hline
  NGC 362   &    -1.52 & 0.22  & - 1.30 & 0.04  &     14.67     &       0.06        &           0.05& 14  \\           
  NGC 1261  &    -1.54 & 0.21  & -1.27 & 0.08 &     16.02     &       0.10        &           0.01& 13  \\           
  NGC 1851  &    -1.57 & 0.30   & -1.18 & 0.08 &     15.42     &       0.08        &           0.02& 21  \\           
  NGC 2419  &    -1.97 & 0.22  & -2.20 & 0.09 &     19.56     &       0.06        &           0.08& 9   \\           
  NGC 3201  &    -1.63 & 0.15  & -1.51 & 0.02 &     13.42     &       0.09        &           0.24& 70  \\           
  NGC 4147  &    -1.61 & 0.04  & -1.78 & 0.08 &     16.33     &       0.06        &           0.02& 6   \\           
  Rup 106   &    -1.80 & 0.13  & -1.78 & 0.08 &     16.73     &       0.35        &           0.20& 13  \\           
  NGC 4590 (M68) &    -2.16 & 0.15  & -2.27 & 0.04 &     15.06     &       0.07        &           0.05& 20  \\           
  NGC 4833  &    -2.08 & 0.11   & -1.89 & 0.05 &     13.98     &       0.09        &           0.32& 9   \\           
  NGC 5024 (M53) &    -2.06 & 0.20   & -2.06 & 0.09 &     16.27     &       0.06        &           0.02& 26  \\           
  NGC 5053  &    -2.05 & 0.18  & -2.30 & 0.08 &     16.14     &       0.07        &           0.01& 8   \\           
  NGC 5139  &    -1.73 & 0.40  & -1.64 & 0.09 &     13.63     &       0.37        &           0.12& 36  \\           
  NGC 5272 (M3)&    -1.71 & 0.24  & -1.50 & 0.05 &     15.03     &       0.09        &           0.01& 23  \\           
  NGC 5466  &    -1.94 & 0.26  & -2.31 & 0.09 &     15.98     &       0.04        &           0   & 13  \\           
  IC 4499   &    -1.74 & 0.22  & -1.62 & 0.09 &     16.44     &       0.07        &           0.23& 61  \\           
  NGC 5824  &    -2.00 & 0.23  & -1.94 & 0.14 &     17.52     &       0.08        &           0.13& 16  \\           
  NGC 5904 (M5) &    -1.57 & 0.28  & -1.33 & 0.02 &     14.33     &       0.09        &           0.03& 38  \\           
  NGC 5986  &    -1.86 & 0.19  & -1.63 & 0.08 &     15.09     &       0.05        &           0.28& 6   \\           
  NGC 6093 (M80) &    -2.05 & 0.31   & -1.75 & 0.08 &     15.00     &       0.33        &           0.18& 6   \\           
  NGC 6121 (M4) &    -1.31 & 0.16  & -1.18 & 0.02 &     11.60     &       0.13        &           0.35& 39  \\           
  NGC 6171 (M107)  &    -1.20 & 0.14    & -1.03 & 0.02  &  14.01     &       0.40        &           0.33& 22  \\           
  NGC 6229  &    -1.51 & 0.25  & -1.43& 0.09 &     17.38     &       0.07        &           0.01& 26  \\           
  NGC 6304  &    -1.58 & 0.50  & -0.37 & 0.07 &     14.63     &       0.25        &           0.54& 6   \\           
  NGC 6341  (M92)&    -2.13 & 0.11  & -2.35 & 0.05 &     14.57     &       0.03        &           0.02& 8   \\           
  NGC 6333 (M9) &    -1.74 & 0.31  & -1.79 & 0.09 &     15.00     &       0.61        &           0.38& 9   \\           
  NGC 6362  &    -1.29 & 0.20  & -1.07 & 0.05 &     14.36     &       0.06        &           0.09& 25  \\           
  NGC 6426  &    -2.10 & 0.29  & - & - &     16.13     &       0.82        &           0.36& 10  \\           
  NGC 6584  &    -1.63 & 0.22  & -1.50 & 0.09  &     15.64     &       0.07        &           0.10& 30  \\           
  NGC 6638  &    -1.37 & 0.20  & -0.99 & 0.07 &     14.77     &       0.23        &           0.41& 6   \\           
  NGC 6642  &    -1.33 & 0.24  & -1.19 & 0.14 &     14.49     &       0.13        &           0.40& 6   \\           
  NGC 6712  &    -1.33 & 0.09  & -1.02 & 0.07 &     14.19     &       0.05        &           0.45& 6   \\           
  NGC 6723  &    -1.34 & 0.28  & -1.10 & 0.07  &     14.59     &       0.07        &           0.05& 9   \\           
  NGC 6864 (M75) &    -1.52 & 0.32  & -1.29 & 0.14 &     16.55     &       0.11        &           0.16& 7   \\           
  NGC 6934  &    -1.68 & 0.20  & -1.56 & 0.09 &     15.99     &       0.06        &           0.10& 47  \\           
  NGC 6981 (M72) &    -1.71 & 0.26  &  -1.48 & 0.07 &     16.15     &       0.12        &           0.05& 33  \\           
  NGC 7006  &    -1.69 & 0.28  & -1.46 & 0.06 &     18.11     &       0.07        &           0.05& 30  \\           
  NGC 7078 (M15) &    -2.19 & 0.26  & -2.33 & 0.02 &     15.08     &       0.08        &           0.10& 44  \\           
  NGC 7089 (M2)&    -1.94 & 0.23  & -1.66 &  0.07 &     15.23     &       0.04        &           0.06& 14  \\           
    \hline
	\end{tabular}
\end{table*}

We have compared the photometric metallicity derived in this study with metallicities estimated from HR spectroscopy, available in the literature \citep{Crestani2021, Pancino2015, Gilligan2021, Nemec2013}. The top panels of Fig.~\ref{fig:comp_lit_hr} show the comparison between the metallicity estimates, while the bottom panels show the residuals ${\rm \Delta [Fe/H] = [Fe/H]_{spec} - [Fe/H]_{phot}}$. The black dashed line represents the one-to-one relation. The mean residual ($\Delta$) and the rms scatter, calculated with respect to zero residuals, are also shown for each sample. There is a good agreement of our photometric metallicity estimates with the HR-spectroscopic metallicities from \citet{Crestani2021} and \citet{Gilligan2021} with small shifts of $\Delta = -0.04$~dex (rms=0.33~dex) and $\Delta = -0.07$~dex (rms=0.39~dex), respectively. There are two strong outliers (WZ Hya and V413 Oph) in the first panel of Fig.~\ref{fig:comp_lit_hr} showing the comparison of our photometric metallicities with values from \citet{Crestani2021} and one strong outlier (WZ Hya) in the third panel, showing the comparison with \citet{Gilligan2021} metallicity estimates. Star V413 Oph ({\it Gaia} DR3 source\_id = 4344296958198114688) was discussed in Section~\ref{subsec:HR}. Its period from {\it Gaia} DR3 catalogue differs from the period provided by \citet{Crestani2021}, thus, its Fourier parameters and, consequently, photometric metallicity could be incorrect. Star WZ Hya was discussed in Section~\ref{subsec:bayes}, as its Fourier parameters provided in {\it Gaia} DR3 catalogue could also be incorrect. Excluding both stars from the comparison, we found negligible shits of $\Delta = -0.02$~dex (rms=0.28~dex) and $\Delta = -0.04$~dex (rms=0.33~dex) between our photometric metallicity estimates and \citet{Crestani2021} and \citet{Gilligan2021} values, respectively. Both samples were used to train our Bayesian predictive model for RRab stars. Agreement between the true and predicted values shows that our model is solid.

At the same time, there are non-negligible shifts between our photometric metallicities and the metallicity values from \citet{Pancino2015} and \citet{Nemec2013}: $\Delta = 0.13$~dex (rms = 0.26~dex) and $\Delta = 0.25$~dex (rms=0.39~dex), respectively, with our values being more metal-poor. The same trend was found by \citet{Crestani2021}, when they made a comparison between their HR-spectroscopic metallicities and those of \citet{Nemec2013} and \citet{Pancino2015} (see Table~4 in \citealt{Crestani2021}). Thus, the shift that we see in Fig~\ref{fig:comp_lit_hr} is mainly related to the shifts between the metallicity scales adopted by \citet{Crestani2021} and those adopted by \citet{Pancino2015} and \citet{Nemec2013}. 

We also compared our photometric metallicities with the metallicities from LR-spectroscopy \citep{Zinn2020, Liu2020}. This comparison is shown in Fig.~\ref{fig:comp_lit_lr}. We found an excellent agreement between our values and those of \citet{Zinn2020} with a shift of $\Delta = 0$~dex and rms of 0.31~dex. The shift between LR-spectroscopic metallicity values from \citet{Liu2020} and our estimates is mild $\Delta = -0.05$~dex (rms=0.35~dex). Still, the right panel of Fig.~\ref{fig:comp_lit_lr} clearly shows that our photometric metallicities are systematically higher than \citet{Liu2020}'s, particularly for RRc stars.
This finding is in agreement with the shift of 0.21~dex between \citet{Crestani2021} and \citet{Liu2020} metallicity values found by \citet{Crestani2021} and discussed in Section~\ref{subsec:bayes}. 

Finally, we compared our [Fe/H] estimates and the photometric metallicities of RRLs calculated using the {\it Gaia} $G$-band photometry by \citet{Iorio2021}, \citet{Dekany2022}, \citet{Li2023} and \citet{Clementini2023}. \citet{Iorio2021} calibrated the $P -\phi_{31}-{\rm [Fe/H]}$ relation using a sample of 84 RRab stars from \citet{Layden1994} with metallicities from low-to-moderate-resolution spectroscopy and period and $\phi_{31}$ Fourier parameter from the {\it Gaia} DR2 catalogue \citep{Clementini2019}. They calibrated the relation for RRc stars using members of GCs assuming as metallicities the spectroscopic values of the corresponding clusters in the \citet{Harris2010} catalogue.
Metal abundances of \citet{Iorio2021} are on the \citet{ZW1984} metallicity scale. To make a comparison with our metallicity estimates which are on the metallicity scale adopted by \citet{Crestani2021}, we firstly transformed the metallicity values from \citet{Iorio2021} to the \citet{Carretta2009} scale using the relation: ${\rm [Fe/H]_{C09}} = 1.105{\rm [Fe/H]_{ZW84}} + 0.160$ \citep{Carretta2009}. We then converted them to the scale adopted by \citet{Crestani2021}, adding a shift of 0.08~dex according to \citet{Crestani2021} and \citet{Mullen2021}. The first two panels of Fig.~\ref{fig:comp_lit_gaia} show the comparison between our metallicity estimates and the metal abundances from \citet{Iorio2021} for RRab (upper panel) and RRc stars (bottom panel), respectively. There is an excellent  agreement between the two metallicity estimates with a mild shift of $\Delta = 0.04$~dex (rms=0.13~dex) for RRc stars, while there is a clear offset of $\Delta = 0.16$~dex (rms=0.19~dex) for RRab stars. Values of metallicity from \citet{Iorio2021} are more metal-rich for low metallicities and more metal-poor for high metallicities. This offset could be due to differences in the training sets used to calibrate the relations (RRLs from \citealt{Crestani2021} in our study versus \citealt{Layden1994} RRL sample in \citealt{Iorio2021}), and differences in period and Fourier parameters of the RRLs ({\it Gaia} DR2 in \citealt{Iorio2021} versus {\it Gaia} DR3 in this study). 

\citet{Dekany2022} used deep learning algorithms to calculate metallicities of $\sim$60,000 RRab stars from the $G$-band light curves published in {\it Gaia} DR2. The metallicity of the training sample of RRLs was estimated using empirical relations derived for the $I$-band photometry, which were calibrated based on the sample of RRLs with HR-spectroscopic estimates of metallicity \citep{Dekany2021}. The training set used by \citet{Dekany2021} consists of RRLs with HR-spectroscopic measurements from \citet{Crestani2021}, \citet{Clementini1995}, \citet{Fernley1996}, \citet{Lambert1996}, \citet{Liu2013}, \citet{Nemec2013}, \citet{Govea2014}, \citet{Pancino2015} and \citet{Andrievsky2018}.  The second panel of Fig.~\ref{fig:comp_lit_gaia} shows the comparison of our photometric [Fe/H] estimates and values from \citet{Dekany2021} for RRab stars. Even though there is a reasonable agreement between the two datasets, there is still  an offset of $\Delta = -0.19$~dex with 0.31~dex scatter. This could be due to the different methods applied to predict metallicity (Bayesian regression versus deep learning algorithms) and differences in the training sets. 

Recently, \citet{Li2023} published $P-\phi_{31}-R_{21}-{\rm [Fe/H]}$ and $P - R_{21} - {\rm [Fe/H]}$ relations for RRab and RRc stars, respectively, calibrated using period and Fourier parameters from the {\it Gaia} DR3 catalogue and metallicities from LR-spectroscopy \citep{Liu2020}. The third panels of Fig.~\ref{fig:comp_lit_gaia} show a comparison of our results with estimates from \citet{Li2023}. For RRab stars, we found a negligible shift of $\Delta = -0.01$~dex (rms=0.11~dex) between our estimates and \citet{Li2023}, with the latter being metallicitites being higher at the metal-rich end and lower than ours at the metal-poor end. 
\citet{Li2023} metallicities for RRc stars are  instead lower than ours ($\Delta=-0.22$~dex, rms=0.29~dex) over the whole metallicity range.
This shift could be due to the offset of 0.21~dex \citep{Crestani2021} between the training sample used to calibrate metallicities in our analysis and the training sample of stars with the LR-spectroscopic measurements from \citet{Liu2020} used by \citet{Li2023} to calibrate their relation (see Section~\ref{subsec:bayes}). It is also worth noting that \citet{Li2023} estimated the metallicities of RRab stars from $P-\phi_{31}-R_{21}-{\rm [Fe/H]}$ relation, where $R_{21} = A_2/A_1$, while we used $P-\phi_{31}-{\rm [Fe/H]}$ relation. At the same time metallicities of RRc stars were estimated by \citet{Li2023} using a $P-R_{21}-{\rm [Fe/H]}$ relation, while a $P-\phi_{31}-A_2-{\rm [Fe/H]}$ relation was applied in our study. This may also have contributed to the larger discrepancy with our estimates, especially, for RRc stars.

Finally, we compared our metallicities with the values provided in the {\it Gaia} DR3 catalogue \citep{Clementini2023}. Results are shown in the fourth panels of Fig.~\ref{fig:comp_lit_gaia}. The metallicity values of RRLs in {\it Gaia} catalogue (both for DR2 and DR3) are derived from quadratic relations published by \citet{Nemec2013}. Those relations are calibrated using 41 RRLs (37 RRab and 4 RRc stars) with metal abundances from HR spectra observed in the field-of-view of the {\it Kepler} space telescope. A sequence of transformations between the Gaia $G$-band and the {\it Kepler} photometric system were necessary to apply \citet{Nemec2013} relations (see details in Sect. 2.1.1 of \citealt{Clementini2019}), which could potentially have caused additional systematic errors. This may explain the discrepancy between metallicity estimates in this study and RRL metallicities in the {\it Gaia} DR3  catalogue seen in Fig.~\ref{fig:comp_lit_gaia}. 


\subsection{Metallicity in GCs}\label{subsec:met_gc}

\begin{figure*}
\includegraphics[trim=120 100 130 30, width = 18cm]{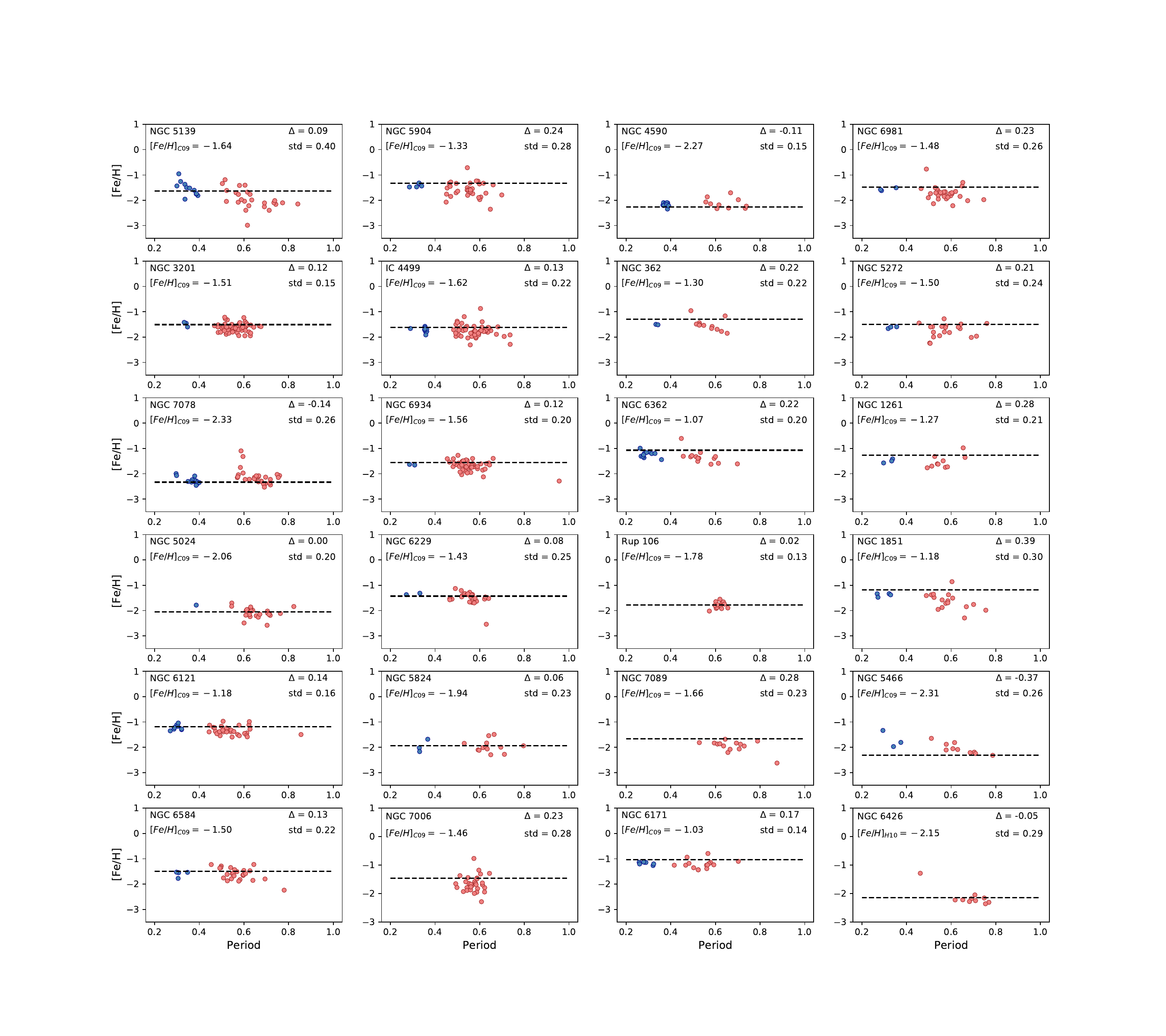}
    \caption{Photometric metallicities of RRLs calculated in this study plotted versus period for RRab (red dots) and RRc (blue dots) stars in 24 GCs of those listed in Table~\ref{tab:gc}, that contain 10 or more RRLs. Black dashed lines show the GC metallicity from \citet{Carretta2009}. The shift between the metallicity of GC from \citet{Carretta2009} and the mean metallicity of RRLs in the GC is labelled ($\Delta$) along with the standard deviation (std) of individual RRL metallicity estimates around the mean value. \citet{Carretta2009} did not provide a metallicity estimate for NGC~6426, thus, we use the metallicity value from \citet{Harris2010} for this cluster.}
    \label{fig:gc}
\end{figure*}

\begin{figure}
\includegraphics[trim = 20 40 40 0, width = 8.5cm]{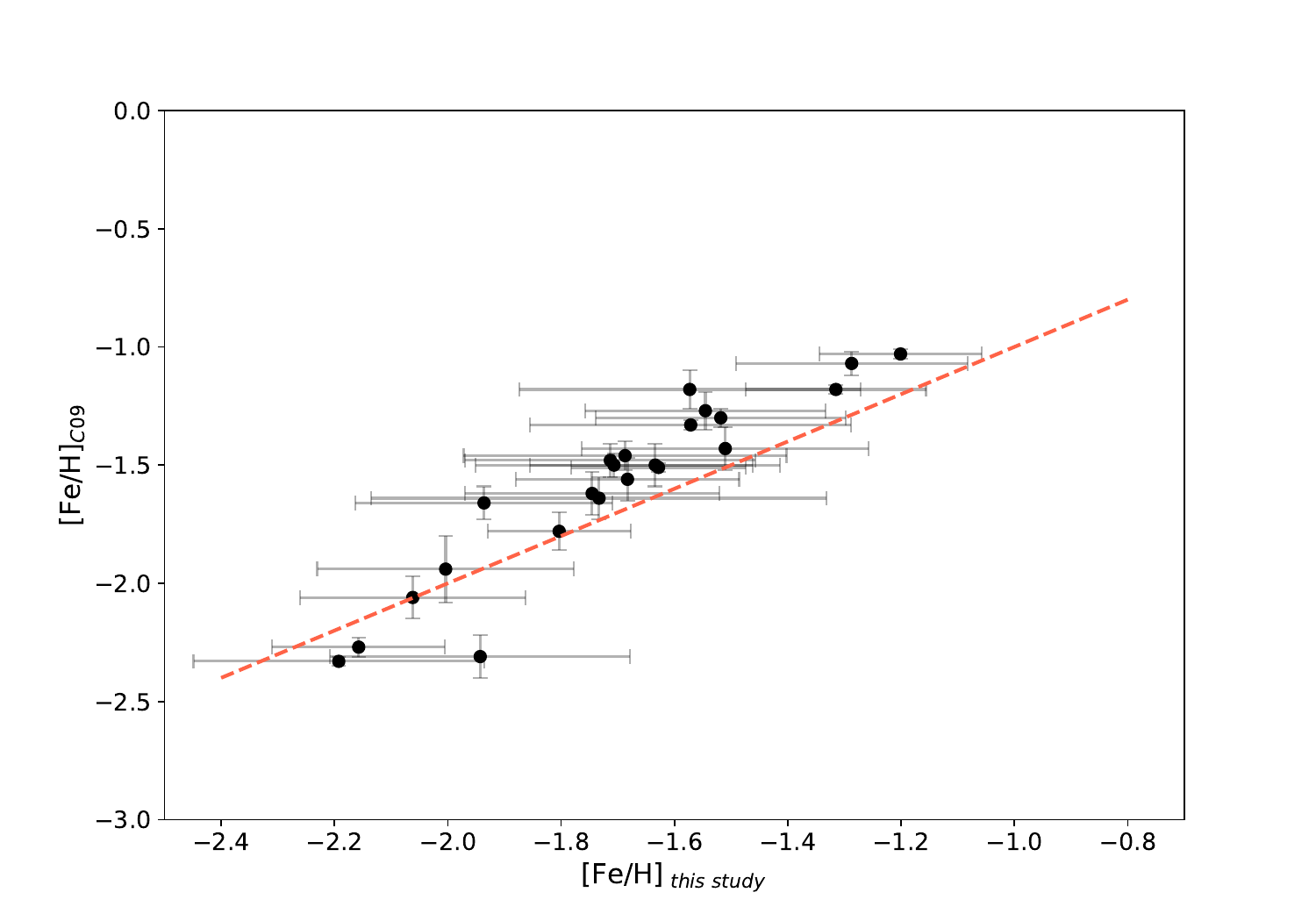}
    \caption{Comparison between the metallicity of GCs from \citet{Carretta2009} and the mean photometric metallicity of RRLs in the GCs calculated in this study. Only GCs containing 10 or more RRLs with photometric metalicity estimates are shown. The red dashed line represents the one-to-one relation.
    }
    \label{fig:gc_C09}
\end{figure}

To further test the quality of our metallicity estimates, we analysed RRLs from our sample belonging to MW GCs whose metal abundance from HR spectra is available in the literature. We used a compilation from the literature to select the member stars of GCs. As a reference compilation of RRLs identified as members of the MW GCs, we have adopted the \citet{Clement2001}'s catalogue\footnote{https://www.astro.utoronto.ca/~cclement/cat/listngc.html}. This catalogue which is constantly updated from the literature, summarizes numbers and types of variable stars in GCs. We restricted our analysis only to GCs with more than five RRLs, for which we estimated a photometric metallicity using Eqs.~\ref{eq:rrab}-\ref{eq:rrc}. As a result, we selected 785 RRLs belonging to 38 GCs (Table~\ref{tab:gc}). Fig.~\ref{fig:gc} shows metallicity versus period for RRab (red dots) and RRc (blue dots) stars in GCs with 10 or more RRLs. We transformed our photometric metallicities that are estimated on the metallicity scale adopted by \citet{Crestani2021} to \citet{Carretta2009} metallicity scale by subtracting 0.08~dex from our values \citep{Crestani2021, Mullen2021}. Black dashed lines represent the \citet{Carretta2009}'s  metallicity for the GCs. Differences between \citet{Carretta2009} values and the mean metallicity we infer from the GC RRLs are labelled ($\Delta$), along with the standard deviation (std) of individual RRL metallicities  from the mean values. \citet{Carretta2009} did not provide metallicity estimate for the GC NGC~6426, thus, we use metallicity value from \citet{Harris2010} for this cluster.

Even though GCs can host different stellar population, metallicity of RRLs in the same GC are expected to have similar values. Fig.~\ref{fig:gc} allows us to check (1) if the scatter in metallicity of RRLs in the same cluster (std) is reasonably small; (2) if there is any dependence of our metallicity values on period, which can hint to some bias in our method; (3) if there is consistency between  [Fe/H]  values of RRab  and RRc stars in the same cluster; (4) if there is consistency between our mean photometric metallicities for RRLs and the GC metallicities from \citet{Carretta2009}. 
We see that apart from some clusters (e.g. NGC 5139, NGC 1851), the scatter of RRLs belonging to the same cluster is relatively small (less than 0.3~dex) and consistent with uncertainties in individual measurements ($\sim0.4$~dex). We do not see any significant dependence of the metallicity on period. Metallicities of RRab and RRc stars in the same cluster are in good agreement. Apart from a few GCs (e.g. NGC 1851, NGC 5466), our metallicity estimates also agree with \citet{Carretta2009} values. In addition, we found that our method can predict metallicities of both metal-poor (e.g. NGC 7078, NGC 5024, NGC 6426, NGC 4590) and metal-rich clusters (e.g. NGC 6121, NGC 6171) with good accuracy.

Fig.~\ref{fig:gc_C09} shows the comparison between metallicities of GCs from \citet{Carretta2009}  and weighted mean of our photometric [Fe/H] estimates of RRLs for each cluster. Only clusters with 10 or more RRLs are shown. Errors in the mean metallicity of RRLs are calculated as the standard deviation of the mean value. As shown in Fig.~\ref{fig:gc_C09} our values are consistent with the metallicities from \citet{Carretta2009} within the errors. 

\section{Distances to RR Lyrae stars}\label{sec:dist}

\begin{figure*}
\includegraphics[width = 17cm]{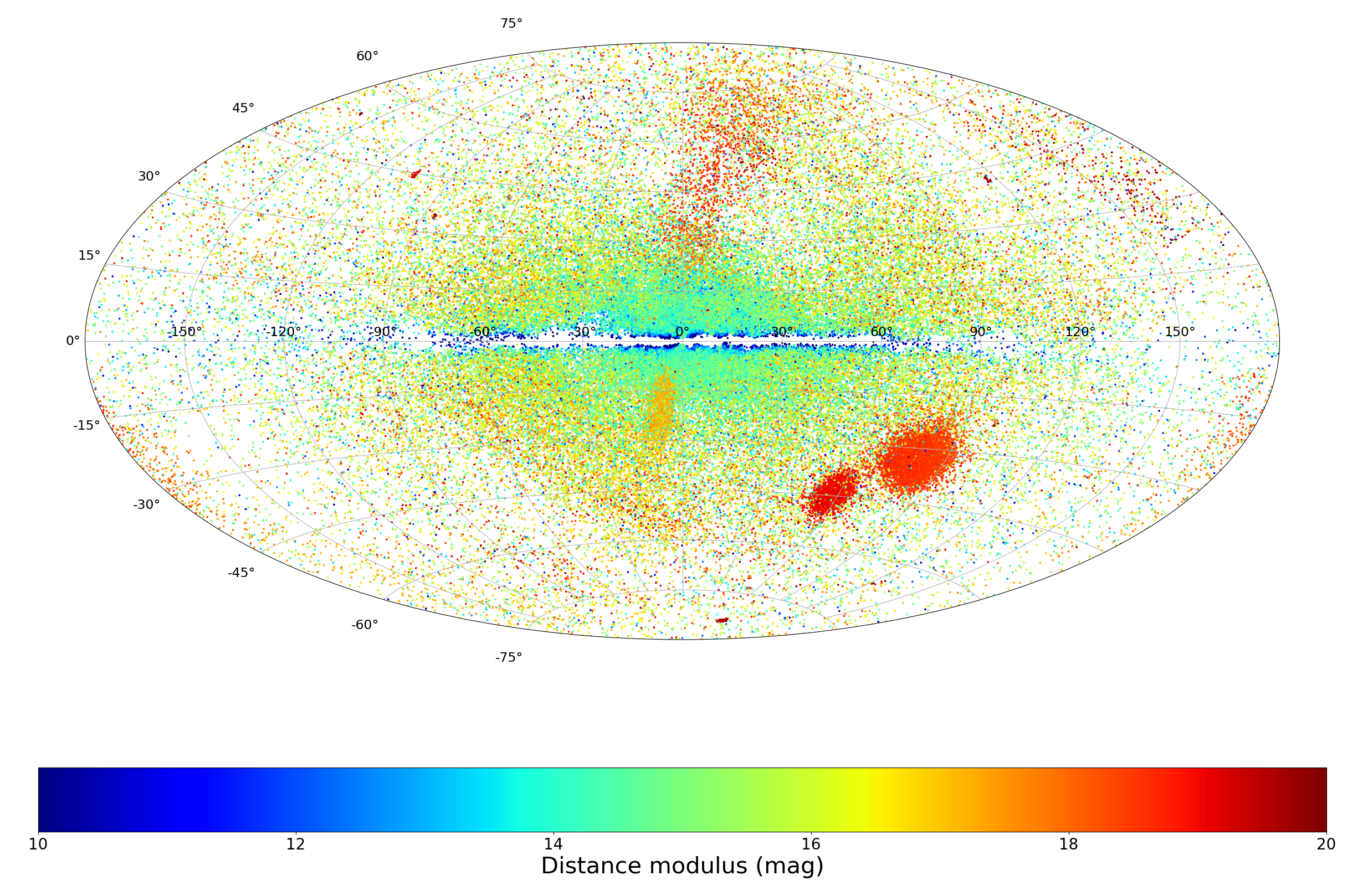}
    \caption{Sky distribution of the 134,642 RRLs with individual distances estimated in this study. Sources are colour-coded by distance modulus.}
    \label{fig:sky_dist}
\end{figure*}

\begin{figure*}
\includegraphics[trim = 90 0 50 20, width = 8.5cm]{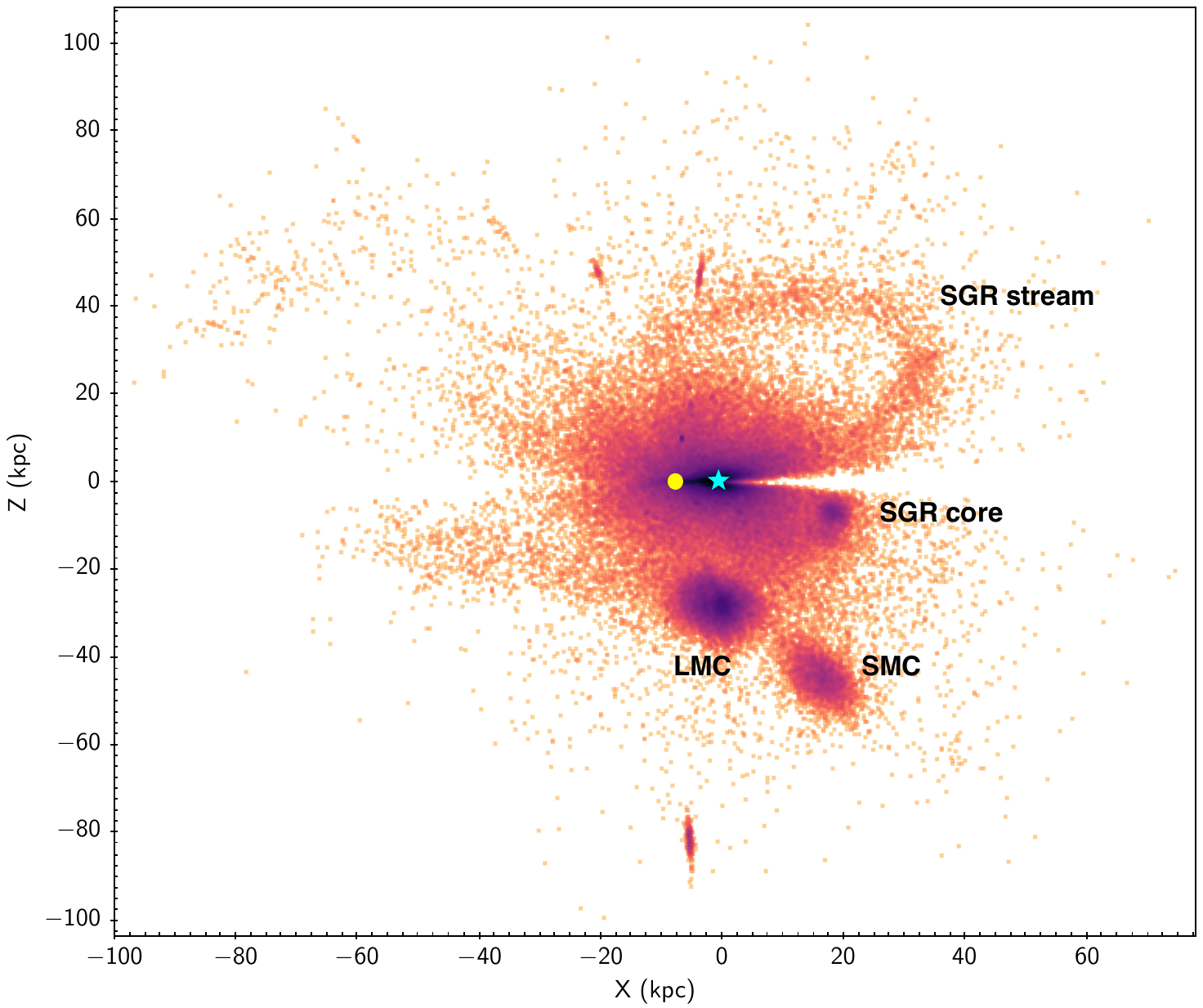}
\includegraphics[trim = 90 0 50 20, width = 8.5cm]{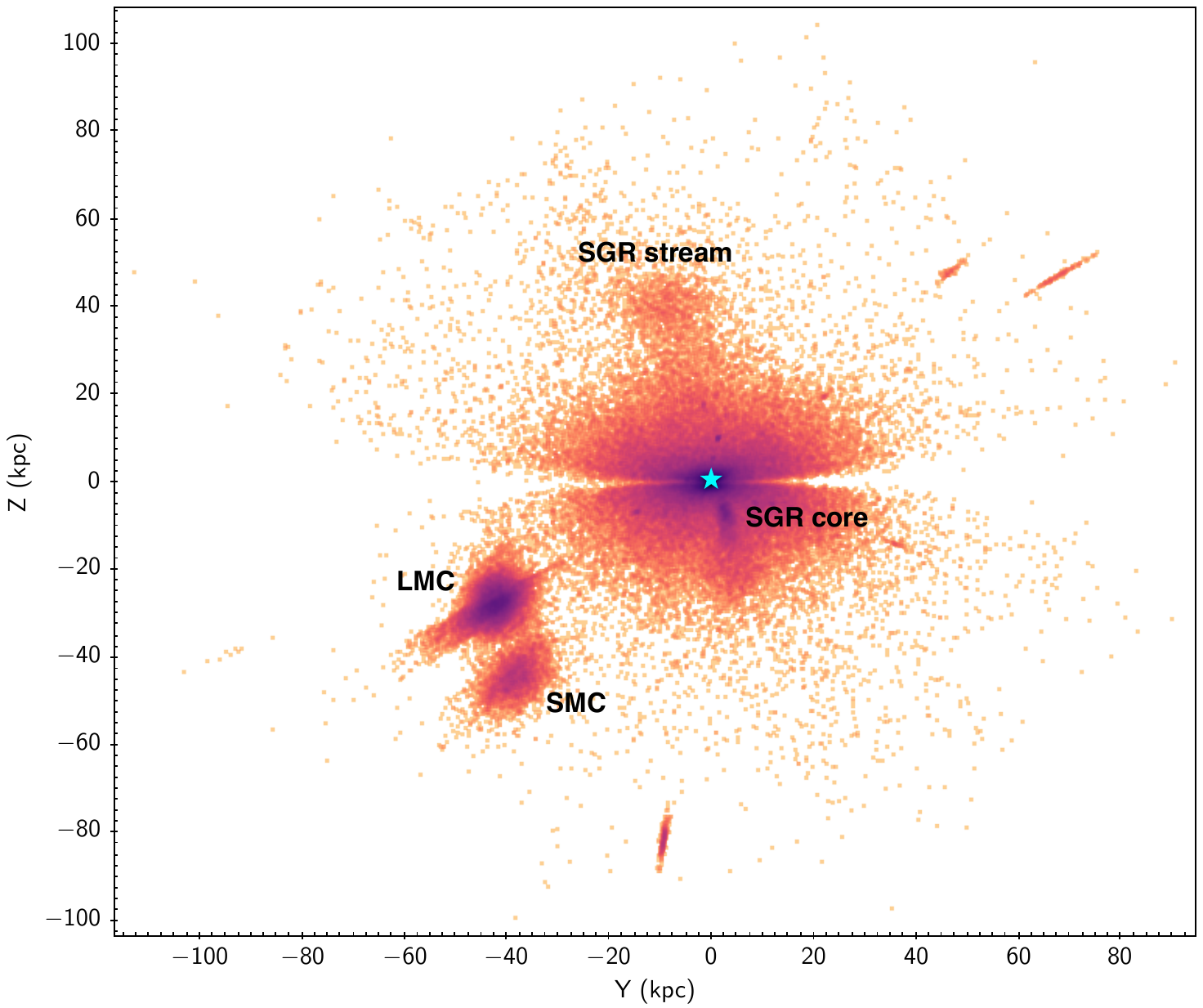}
\caption{Spatial distribution of RRLs in our sample on the Cartesian X-Z ({\it left panel}) and Y-Z ({\it right panel}) planes. The blue star and the yellow circle show the centre of our Galaxy and the Sun, respectively. The LMC, SMC, the Sgr stream and core are clearly seen.}
    \label{fig:xyz}
\end{figure*}

We used the metallicity values derived in the previous sections to calculate individual distances to the RRLs in our sample. Following the approach of \citet{Li2023}, we adopted ${\rm E(B-V)}$ reddening values from \citet{sfd1998} for the majority of stars. For RRLs located in the Magellanic Clouds, we adopted instead values of $E(V-I)$ from \citet{Skowron2021}, which we transformed to ${\rm E(B-V)}$ using the relation ${\rm E(B-V) = E(V-I)/1.318}$ from \citet{Skowron2021}. Finally, we used reddening values from \citet{Harris2010} for RRLs in GCs. In this way, we were able to find reddening values for 134,665 stars, for which we have photometric metallicities estimated with Eqs.~\ref{eq:rrab}-\ref{eq:rrc}. Uncertainties of the reddening values were taken from the corresponding studies apart from \citet{Harris2010} values, for which we adopted a reddening uncertainty of 0.05~mag. Individual extinction values were calculated using the total-to-selective extinction ratio $R_G = 2.516\pm0.036$ \citep{Huang2021}.

 Distances were calculated using the $M_G - {\rm [Fe/H]}$ relation by \citealt{Garofalo2022} (their Eq. 18), which was calibrated using the hierarchical Bayesian approach and accurate parallaxes of bright MW RRLs published in the {\it Gaia} EDR3 catalogue. \citet{Garofalo2022} used metallicities of GCs on the \citet{Carretta2009} metallicity scale. Thus, we transformed our metallicity values to the \citet{Carretta2009} scale by subtracting 0.08~dex. For a handful of stars (23, 0.0002\% of the sample) in the central region of the disk, the obtained distance moduli have non-physical negative values due to significantly overestimated reddening. We discarded these stars, thus remaining with a sample of 134,642 RRLs with individual distance estimates. 

Uncertainties in our distance estimates come from  uncertainties in extinction and individual metallicity values (see Section~\ref{sec:met_val}), uncertainties in the coefficients of the $M_G - {\rm [Fe/H]}$ relation, and intrinsic scatter of the fit ($\sigma$ = 0.13). We estimated the uncertainties in distance moduli using a Monte Carlo simulation approach. For each star, 1000 iterations were performed. The random values were sampled from the error distributions of extinction, metallicity and coefficients, while also simulating the intrinsic dispersion of the $M_G - {\rm [Fe/H]}$ relation ($\sigma$). Collecting the distance values obtained from the Monte Carlo simulation allowed us to estimate uncertainties in distance moduli.

The sky distribution of the 134,642 RRLs in our sample, colour-coded by distance, is shown in Fig.~\ref{fig:sky_dist}. The closest stars to us are located in the disk of the MW, as expected. The Sagittarius (Sgr) stream and core are clearly seen. In Table~\ref{tab:gc} we report mean value and standard deviation of the GC’s distance
moduli, calculated using individual distance estimates to the RRLs in each GC.

Finally, we calculated the 3-D positions of the RRLs in our sample from their coordinates and estimated distances. The position of the Sun was assumed to be on the X-axis of the right-handed system. The X-axis points from the position of the Sun to the Galactic centre, while the Y-axis points towards Galactic longitude $l=90^{\circ}$. The Z-axis points towards the North Galactic Pole ($b = 90^{\circ}$). The Sun was assumed to be at a distance of 8.122~kpc from the Galactic centre \citep{Gravity2018}. Fig.~\ref{fig:xyz} shows the distribution of RRLs on the X-Z (left panel) and Y-Z (right panel) planes. Well-known structures, such as the Large Magellanic Cloud (LMC), Small Magellanic Cloud (SMC), the Sgr stream and the core of the Sgr dwarf spheroidal galaxy are clearly seen. This shows the potential of distance and metallicity measurements of RRLs obtained in this work for studying substructures, located in and outside of our Galaxy.

\section{The Magellanic Clouds}\label{sec:mc}

\begin{figure}
\includegraphics[trim = 30 60 30 40, width=\columnwidth]{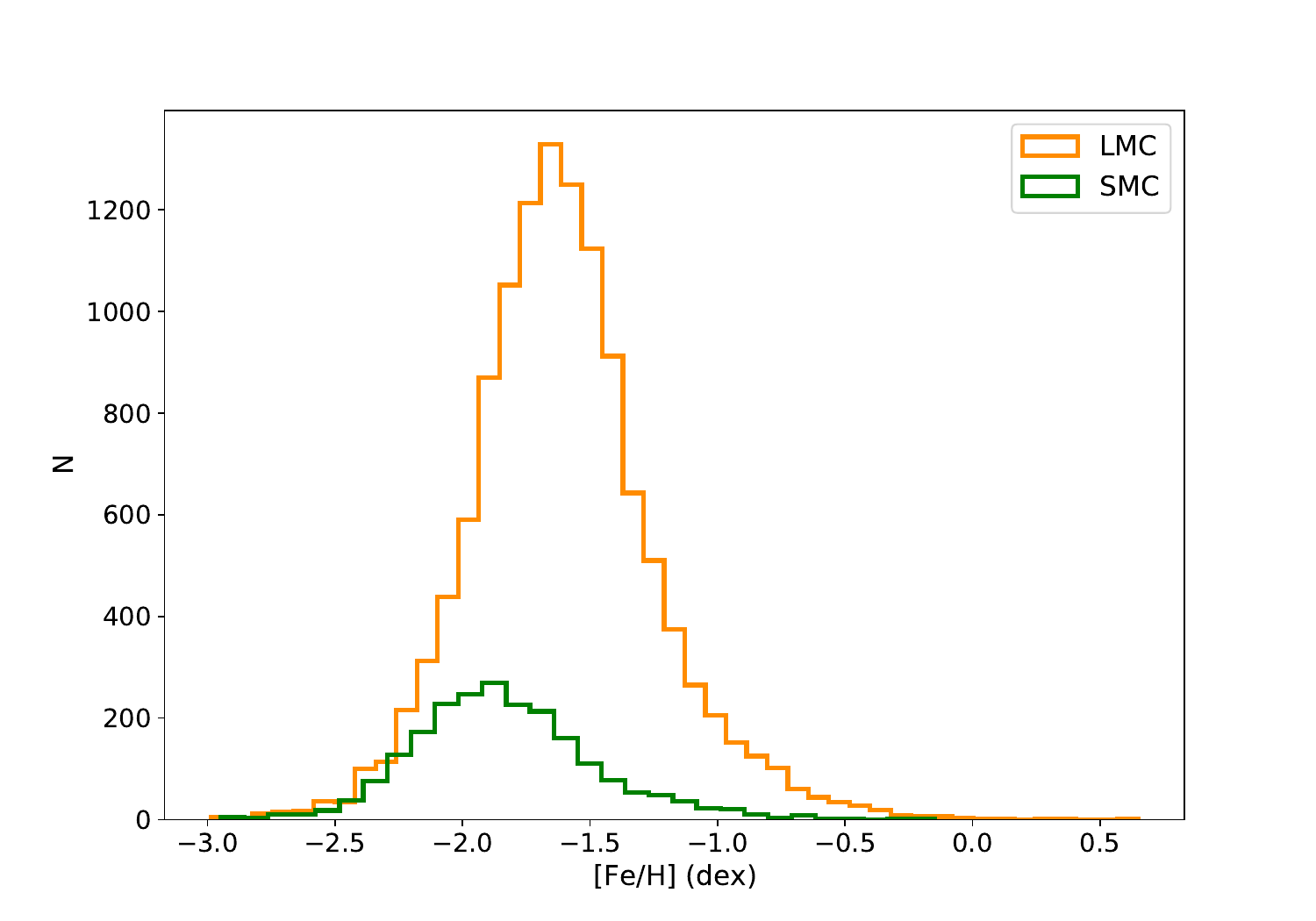}
\caption{Metallicity distribution of the RRLs in the LMC and SMC.}\label{fig:hist_mc}
\end{figure}

\begin{figure*}
\includegraphics[trim = 0 0 0 0, width=16cm]{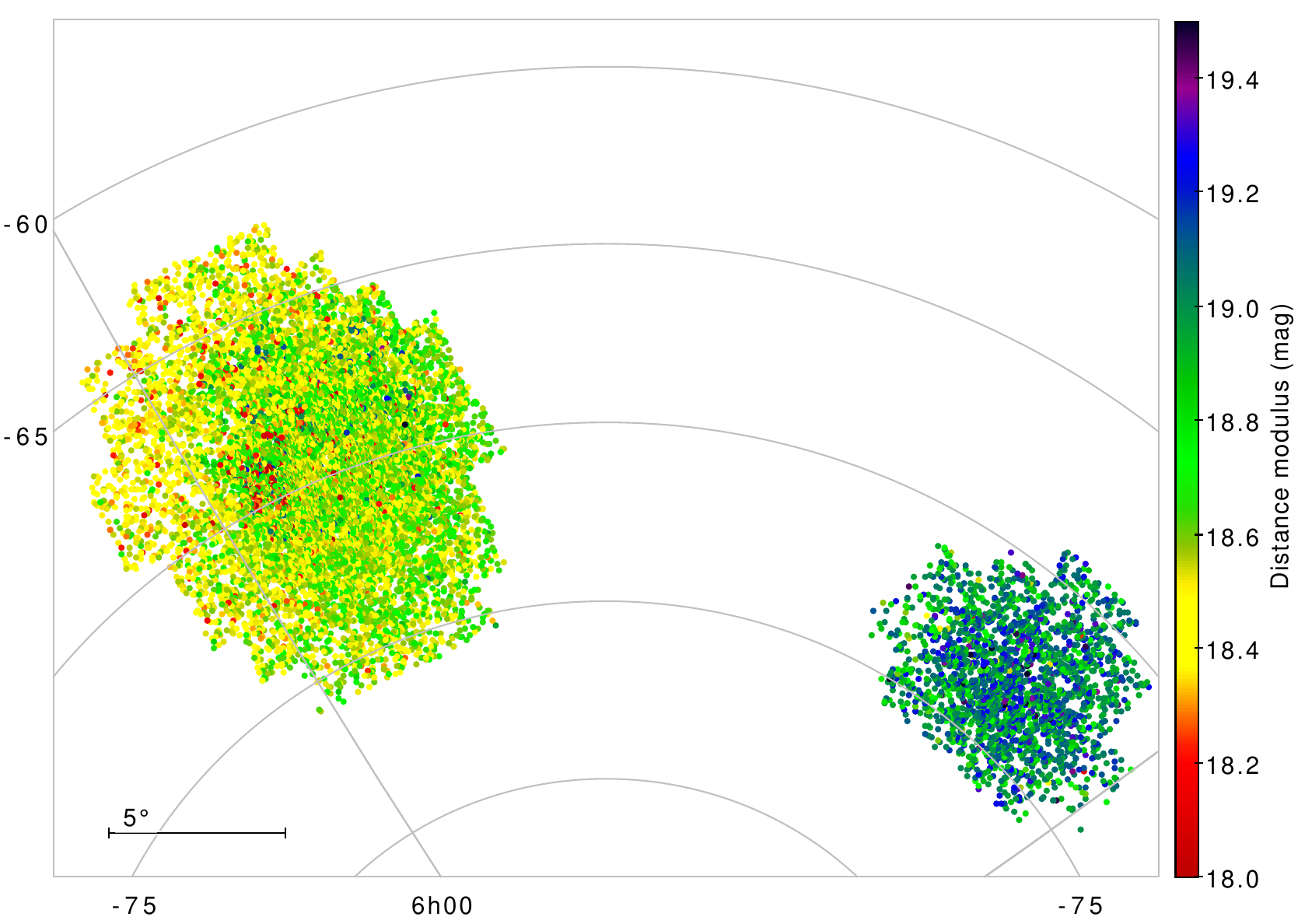}
\caption{Sky distribution of the RRLs in the LMC and SMC, colour-coded by distance modulus}\label{fig:mc_sky}
\end{figure*}

To evaluate the quality of our distance and metallicity estimates, we analyse RRLs located in the LMC and SMC, the biggest MW satellites traditionally used as laboratories for studying different stellar populations. \citet{Cusano2021} analysed a sample of $\sim$ 22,000 RRLs located in the LMC, whose membership was confirmed by means of the $PL$ relation in the $K_{\rm s}$ band. We cross-matched this catalogue against our sample of RRLs with photometric metallicity and distance estimates and found 12,239 stars in common. The metallicity distribution of these LMC RRLs is shown by the orange histogram in Fig.~\ref{fig:hist_mc}. The weighted mean photometric metallicity of the RRLs in the LMC is ${\rm [Fe/H]_{LMC} = -1.63\pm0.36}$~dex, where the uncertainty is calculated as the standard deviation of the mean value. This metallicity value is in good agreement with ${\rm [Fe/H]_{LMC} = -1.48\pm0.03\pm0.06}$~dex measured by \citet{Gratton2004} from  low-resolution spectra of 98 RRLs in the bar of the LMC and 
${\rm [Fe/H]_{LMC} = -1.53\pm0.02}$~dex reported by \citet{Borissova2006} based on the analysis of 100 RRLs in the LMC. \citet{Skowron2016} obtained the median value of photometric metallicities of RRLs in the LMC ${\rm [Fe/H]_{ZW84} = -1.59\pm0.31}$ on the \citet{ZW1984} metallicity scale that once transformed to the metallicity scale adopted by \citet{Crestani2021} used in the present study (see Sect.~\ref{subsec:comp_lit}) provides the value ${\rm [Fe/H]_{C21} = -1.52}$~dex in good agreement with our estimate. We also measured a mean distance modulus for the RRLs in the LMC: $\mu_{\rm LMC}=18.55\pm0.18$~mag, in good agreement within the errors with previous measurements based on Cepheids ($\mu_{\rm LMC}=18.477\pm0.033$~mag, \citealt{Freedman2012}), RRLs ($\mu_{\rm LMC}=18.50\pm0.16$~mag, \citealt{Muraveva2018b}) and eclipsing binaries ($\mu_{\rm LMC}=18.48\pm0.03$~mag, \citealt{Pietrzynski2019}).

In \citet{Muraveva2018c}, we analysed 2997 fundamental mode RRLs, which were confirmed to belong to the SMC based on their position on the $PL$ relation in the $K_{\rm s}$ band. We cross-matched this catalogue against our sample of RRLs with photometric metallicity and distance estimates and found 2203 stars in common. Their metallicity distribution is shown by the green histogram in Fig.~\ref{fig:hist_mc}. The weighted mean metallicity of the RRLs in the SMC is ${\rm [Fe/H]_{SMC}=-1.86\pm0.36}$~dex in perfect agreement with the value ${\rm [Fe/H]_{C21}=-1.80}$~dex, obtained by transforming to \citet{Crestani2021} metallicity scale the value ${\rm [Fe/H]_{SMC}=-1.85\pm0.35}$~dex on the \citet{ZW1984} scale derived by \citet{Skowron2016}.
The mean distance modulus of the SMC from RRLs is $\mu_{\rm SMC}=19.01\pm 0.17$~mag in good agreement within the errors with previous values based on Cepheids ($\mu_{\rm SMC}=18.96\pm 0.01\pm0.03$~mag, \citealt{Skowcroft2016}) and eclipsing binaries ($\mu_{\rm SMC}=18.977\pm 0.016\pm0.028$~mag, \citealt{Graczyk2020}).

Fig.~\ref{fig:mc_sky} shows the sky distribution of the RRLs in the LMC and SMC with the colour encoding the individual distance moduli of the stars. It is clearly seen that, as expected, the SMC is located farther from us than the LMC. Moreover, the north-eastern part of the LMC, as traced by RRLs, is closer to us than the south-western part, which has already been reported in the literature (e.g. \citealt{Cusano2021}). This once again confirms the reliability of our distance estimates.


\section{Summary}\label{sec:summ}

We have obtaied new $P -\phi_{31}-{\rm [Fe/H]}$ and $P -\phi_{31}- A_2 - {\rm [Fe/H]}$ relations for RRab and RRc stars, respectively. The relations are based on the periods and Fourier parameters of the {\it Gaia} $G$-band light curves of RRLs published in {\it Gaia} DR3 and spectroscopic metallicites of RRLs available in the literature. Both relations are calibrated on the metallicity scale adopted by \citet{Crestani2021}. We applied a feature selection procedure based on the cross-validation to choose the parameters most relevant for metallicity determination. The relations were derived using the Bayesian fitting approach which allowed us to carefully take into account uncertainties in the parameters and the intrinsic scatter of the fit. The RMSE of the predicted metallicity values for the RRLs in the training sample are 0.28~dex and 0.21~dex for RRab and RRc stars, respectively, comparable with the typical uncertainty of metallicity measurements from LR spectroscopy. 

We applied the newly derived relations to estimate the metallicity of 134,769 RRLs from the {\it Gaia} DR3 catalogue. We compared our metallicities with metallicity estimates from HR- and LR-spectroscopy and with photometric metallicities available in the literature that were derived from {\it Gaia} $G$-band light curves of RRLs.
We found a good agreement (within 0.25~dex) between our values and the metallicity estimates from the literature for most studies. We also analysed the metallicity of RRLs in 38 MW GCs and found that the photometric metallicities of RRLs located in the same cluster have scatter less than 0.3~dex for the majority of GCs, do not show a residual dependence on the period, and that there is consistency between metallicities of RRab and RRc stars in the same cluster. We provided new estimates of metallicities for the 38 GCs based on their RRLs and found that they they are in agreement with the values from \citet{Carretta2009} within the errors. 

We used our metallicity measurements to calculate the distances to 134,642 RRLs employing the $M_G - {\rm [Fe/H]}$ relation from \citet{Garofalo2022}, which is calibrated on {\it Gaia} EDR3 parallaxes. We used these distances to estimate the mean distances to 38 MW GCs and to map the LMC, the SMC, the Sgr stream and its core as traced by RRLs on the Cartesian coordinates plane. 
We calculated mean metallicity and distance to the LMC: ${\rm [Fe/H]_{LMC} = -1.63\pm0.36}$~dex and $\mu_{\rm LMC}=18.55\pm0.18$~mag, and the SMC: ${\rm [Fe/H]_{SMC}=-1.86\pm0.36}$~dex and $\mu_{\rm SMC}=19.01\pm 0.17$~mag, in excellent agreement with previous estimates in the literature. We also confirmed that the RRLs in the north-eastern part of the LMC are closer to us than RRLs in the south-western part \citep{Cusano2021}.
 Our results show that the catalogue of $\sim$134,000 RRLs with distances and metallicities provided in this study is a powerful tool to study the structure and chemical abundance of the MW and the Local Group galaxies and to search for the new substructures in our Galaxy and beyond.

\section*{ACKNOWLEDGEMENTS}

This work uses data from the European Space Agency mission {\it Gaia} (https://www.cosmos.esa.int/gaia), processed by the {\it Gaia} Data Processing and Analysis Consortium (DPAC; https: //www.cosmos.esa.int/web/gaia/dpac/consortium). Funding for the DPAC has been provided by national institutions, in particular the institutions participating in the {\it Gaia} Multilateral Agreement. Support to this study has been provided by INAF Mini-Grant (PI: Tatiana Muraveva), by the Agenzia Spaziale Italiana (ASI) through contract and ASI 2018-24-HH.0, and by Premiale 2015, MIning The Cosmos - Big Data and Innovative Italian Technology for Frontiers Astrophysics and Cosmology (MITiC; P.I.B.Garilli).
\section*{Data Availability}

 The data underlying this article are available in the article and in its online supplementary material.



\bibliographystyle{mnras}
\bibliography{draft.bbl} 

\begin{thebibliography}{}
\makeatletter
\relax
\def\mn@urlcharsother{\let\do\@makeother \do\$\do\&\do\#\do\^\do\_\do\%\do\~}
\def\mn@doi{\begingroup\mn@urlcharsother \@ifnextchar [ {\mn@doi@}
  {\mn@doi@[]}}
\def\mn@doi@[#1]#2{\def\@tempa{#1}\ifx\@tempa\@empty \href
  {http://dx.doi.org/#2} {doi:#2}\else \href {http://dx.doi.org/#2} {#1}\fi
  \endgroup}
\def\mn@eprint#1#2{\mn@eprint@#1:#2::\@nil}
\def\mn@eprint@arXiv#1{\href {http://arxiv.org/abs/#1} {{\tt arXiv:#1}}}
\def\mn@eprint@dblp#1{\href {http://dblp.uni-trier.de/rec/bibtex/#1.xml}
  {dblp:#1}}
\def\mn@eprint@#1:#2:#3:#4\@nil{\def\@tempa {#1}\def\@tempb {#2}\def\@tempc
  {#3}\ifx \@tempc \@empty \let \@tempc \@tempb \let \@tempb \@tempa \fi \ifx
  \@tempb \@empty \def\@tempb {arXiv}\fi \@ifundefined
  {mn@eprint@\@tempb}{\@tempb:\@tempc}{\expandafter \expandafter \csname
  mn@eprint@\@tempb\endcsname \expandafter{\@tempc}}}

\bibitem[\protect\citeauthoryear{{Andrievsky} et~al.,}{{Andrievsky}
  et~al.}{2018}]{Andrievsky2018}
{Andrievsky} S.,  et~al., 2018, \mn@doi [\pasp] {10.1088/1538-3873/aa9783},
  \href {https://ui.adsabs.harvard.edu/abs/2018PASP..130b4201A} {130, 024201}

\bibitem[\protect\citeauthoryear{{Belokurov}, {Deason}, {Koposov}, {Catelan},
  {Erkal}, {Drake}  \& {Evans}}{{Belokurov} et~al.}{2018}]{Belokurov2018}
{Belokurov} V.,  {Deason} A.~J.,  {Koposov} S.~E.,  {Catelan} M.,  {Erkal} D.,
  {Drake} A.~J.,   {Evans} N.~W.,  2018, \mn@doi [\mnras]
  {10.1093/mnras/sty615}, \href
  {https://ui.adsabs.harvard.edu/abs/2018MNRAS.477.1472B} {477, 1472}

\bibitem[\protect\citeauthoryear{{Bono}, {Caputo}, {Castellani}, {Marconi},
  {Storm}  \& {Degl'Innocenti}}{{Bono} et~al.}{2003}]{Bono2003}
{Bono} G.,  {Caputo} F.,  {Castellani} V.,  {Marconi} M.,  {Storm} J.,
  {Degl'Innocenti} S.,  2003, \mn@doi [\mnras]
  {10.1046/j.1365-8711.2003.06878.x}, \href
  {https://ui.adsabs.harvard.edu/abs/2003MNRAS.344.1097B} {344, 1097}

\bibitem[\protect\citeauthoryear{{Borissova}, {Minniti}, {Rejkuba}  \&
  {Alves}}{{Borissova} et~al.}{2006}]{Borissova2006}
{Borissova} J.,  {Minniti} D.,  {Rejkuba} M.,   {Alves} D.,  2006, \mn@doi
  [\aap] {10.1051/0004-6361:20054132}, \href
  {https://ui.adsabs.harvard.edu/abs/2006A&A...460..459B} {460, 459}

\bibitem[\protect\citeauthoryear{{Carretta}, {Bragaglia}, {Gratton}, {D'Orazi}
  \& {Lucatello}}{{Carretta} et~al.}{2009}]{Carretta2009}
{Carretta} E.,  {Bragaglia} A.,  {Gratton} R.,  {D'Orazi} V.,   {Lucatello} S.,
   2009, \mn@doi [\aap] {10.1051/0004-6361/200913003}, \href
  {https://ui.adsabs.harvard.edu/abs/2009A&A...508..695C} {508, 695}

\bibitem[\protect\citeauthoryear{{Chadid}, {Sneden}  \& {Preston}}{{Chadid}
  et~al.}{2017}]{Chadid2017}
{Chadid} M.,  {Sneden} C.,   {Preston} G.~W.,  2017, \mn@doi [\apj]
  {10.3847/1538-4357/835/2/187}, \href
  {https://ui.adsabs.harvard.edu/abs/2017ApJ...835..187C} {835, 187}

\bibitem[\protect\citeauthoryear{{Clement} et~al.,}{{Clement}
  et~al.}{2001}]{Clement2001}
{Clement} C.~M.,  et~al., 2001, \mn@doi [\aj] {10.1086/323719}, \href
  {https://ui.adsabs.harvard.edu/abs/2001AJ....122.2587C} {122, 2587}

\bibitem[\protect\citeauthoryear{{Clementini}, {Carretta}, {Gratton},
  {Merighi}, {Mould}  \& {McCarthy}}{{Clementini}
  et~al.}{1995}]{Clementini1995}
{Clementini} G.,  {Carretta} E.,  {Gratton} R.,  {Merighi} R.,  {Mould} J.~R.,
   {McCarthy} J.~K.,  1995, \mn@doi [\aj] {10.1086/117692}, \href
  {https://ui.adsabs.harvard.edu/abs/1995AJ....110.2319C} {110, 2319}

\bibitem[\protect\citeauthoryear{{Clementini}, {Gratton}, {Bragaglia},
  {Carretta}, {Di Fabrizio}  \& {Maio}}{{Clementini}
  et~al.}{2003}]{Clementini2003}
{Clementini} G.,  {Gratton} R.,  {Bragaglia} A.,  {Carretta} E.,  {Di Fabrizio}
  L.,   {Maio} M.,  2003, \mn@doi [\aj] {10.1086/367773}, \href
  {https://ui.adsabs.harvard.edu/abs/2003AJ....125.1309C} {125, 1309}

\bibitem[\protect\citeauthoryear{{Clementini} et~al.,}{{Clementini}
  et~al.}{2019}]{Clementini2019}
{Clementini} G.,  et~al., 2019, \mn@doi [\aap] {10.1051/0004-6361/201833374},
  \href {https://ui.adsabs.harvard.edu/abs/2019A&A...622A..60C} {622, A60}

\bibitem[\protect\citeauthoryear{{Clementini} et~al.,}{{Clementini}
  et~al.}{2023}]{Clementini2023}
{Clementini} G.,  et~al., 2023, \mn@doi [\aap] {10.1051/0004-6361/202243964},
  \href {https://ui.adsabs.harvard.edu/abs/2023A&A...674A..18C} {674, A18}

\bibitem[\protect\citeauthoryear{{Crestani} et~al.,}{{Crestani}
  et~al.}{2021}]{Crestani2021}
{Crestani} J.,  et~al., 2021, \mn@doi [\apj] {10.3847/1538-4357/abd183}, \href
  {https://ui.adsabs.harvard.edu/abs/2021ApJ...908...20C} {908, 20}

\bibitem[\protect\citeauthoryear{{Cusano} et~al.,}{{Cusano}
  et~al.}{2021}]{Cusano2021}
{Cusano} F.,  et~al., 2021, \mn@doi [\mnras] {10.1093/mnras/stab901}, \href
  {https://ui.adsabs.harvard.edu/abs/2021MNRAS.504....1C} {504, 1}

\bibitem[\protect\citeauthoryear{{D{\'e}k{\'a}ny} \& {Grebel}}{{D{\'e}k{\'a}ny}
  \& {Grebel}}{2022}]{Dekany2022}
{D{\'e}k{\'a}ny} I.,  {Grebel} E.~K.,  2022, \mn@doi [\apjs]
  {10.3847/1538-4365/ac74ba}, \href
  {https://ui.adsabs.harvard.edu/abs/2022ApJS..261...33D} {261, 33}

\bibitem[\protect\citeauthoryear{{D{\'e}k{\'a}ny}, {Grebel}  \&
  {Pojma{\'n}ski}}{{D{\'e}k{\'a}ny} et~al.}{2021}]{Dekany2021}
{D{\'e}k{\'a}ny} I.,  {Grebel} E.~K.,   {Pojma{\'n}ski} G.,  2021, \mn@doi
  [\apj] {10.3847/1538-4357/ac106f}, \href
  {https://ui.adsabs.harvard.edu/abs/2021ApJ...920...33D} {920, 33}

\bibitem[\protect\citeauthoryear{{Drake} et~al.,}{{Drake}
  et~al.}{2013}]{Drake2013}
{Drake} A.~J.,  et~al., 2013, \mn@doi [\apj] {10.1088/0004-637X/763/1/32},
  \href {https://ui.adsabs.harvard.edu/abs/2013ApJ...763...32D} {763, 32}

\bibitem[\protect\citeauthoryear{{Fernley} \& {Barnes}}{{Fernley} \&
  {Barnes}}{1996}]{Fernley1996}
{Fernley} J.,  {Barnes} T.~G.,  1996, \aap, \href
  {https://ui.adsabs.harvard.edu/abs/1996A&A...312..957F} {312, 957}

\bibitem[\protect\citeauthoryear{{For}, {Sneden}  \& {Preston}}{{For}
  et~al.}{2011}]{For2011}
{For} B.-Q.,  {Sneden} C.,   {Preston} G.~W.,  2011, \mn@doi [\apjs]
  {10.1088/0067-0049/197/2/29}, \href
  {https://ui.adsabs.harvard.edu/abs/2011ApJS..197...29F} {197, 29}

\bibitem[\protect\citeauthoryear{{Freedman}, {Madore}, {Scowcroft}, {Burns},
  {Monson}, {Persson}, {Seibert}  \& {Rigby}}{{Freedman}
  et~al.}{2012}]{Freedman2012}
{Freedman} W.~L.,  {Madore} B.~F.,  {Scowcroft} V.,  {Burns} C.,  {Monson} A.,
  {Persson} S.~E.,  {Seibert} M.,   {Rigby} J.,  2012, \mn@doi [\apj]
  {10.1088/0004-637X/758/1/24}, \href
  {https://ui.adsabs.harvard.edu/abs/2012ApJ...758...24F} {758, 24}

\bibitem[\protect\citeauthoryear{{GRAVITY Collaboration} et~al.,}{{GRAVITY
  Collaboration} et~al.}{2018}]{Gravity2018}
{GRAVITY Collaboration} et~al., 2018, \mn@doi [\aap]
  {10.1051/0004-6361/201833718}, \href
  {https://ui.adsabs.harvard.edu/abs/2018A&A...615L..15G} {615, L15}

\bibitem[\protect\citeauthoryear{{Gaia Collaboration} et~al.,}{{Gaia
  Collaboration} et~al.}{2016}]{Prusti2016}
{Gaia Collaboration} et~al., 2016, \mn@doi [\aap]
  {10.1051/0004-6361/201629272}, \href
  {https://ui.adsabs.harvard.edu/abs/2016A&A...595A...1G} {595, A1}

\bibitem[\protect\citeauthoryear{{Gaia Collaboration} et~al.,}{{Gaia
  Collaboration} et~al.}{2021}]{Brown2021}
{Gaia Collaboration} et~al., 2021, \mn@doi [\aap]
  {10.1051/0004-6361/202039657}, \href
  {https://ui.adsabs.harvard.edu/abs/2021A&A...649A...1G} {649, A1}

\bibitem[\protect\citeauthoryear{{Gaia Collaboration} et~al.,}{{Gaia
  Collaboration} et~al.}{2023}]{Vallenari2023}
{Gaia Collaboration} et~al., 2023, \mn@doi [\aap]
  {10.1051/0004-6361/202243940}, \href
  {https://ui.adsabs.harvard.edu/abs/2023A&A...674A...1G} {674, A1}

\bibitem[\protect\citeauthoryear{{Garofalo}, {Tantalo}, {Cusano}, {Clementini},
  {Calura}, {Muraveva}, {Paris}  \& {Speziali}}{{Garofalo}
  et~al.}{2021}]{Garofalo2021}
{Garofalo} A.,  {Tantalo} M.,  {Cusano} F.,  {Clementini} G.,  {Calura} F.,
  {Muraveva} T.,  {Paris} D.,   {Speziali} R.,  2021, \mn@doi [\apj]
  {10.3847/1538-4357/ac0253}, \href
  {https://ui.adsabs.harvard.edu/abs/2021ApJ...916...10G} {916, 10}

\bibitem[\protect\citeauthoryear{{Garofalo}, {Delgado}, {Sarro}, {Clementini},
  {Muraveva}, {Marconi}  \& {Ripepi}}{{Garofalo} et~al.}{2022}]{Garofalo2022}
{Garofalo} A.,  {Delgado} H.~E.,  {Sarro} L.~M.,  {Clementini} G.,  {Muraveva}
  T.,  {Marconi} M.,   {Ripepi} V.,  2022, \mn@doi [\mnras]
  {10.1093/mnras/stac735}, \href
  {https://ui.adsabs.harvard.edu/abs/2022MNRAS.513..788G} {513, 788}

\bibitem[\protect\citeauthoryear{{Gilligan} et~al.,}{{Gilligan}
  et~al.}{2021}]{Gilligan2021}
{Gilligan} C.~K.,  et~al., 2021, \mn@doi [\mnras] {10.1093/mnras/stab857},
  \href {https://ui.adsabs.harvard.edu/abs/2021MNRAS.503.4719G} {503, 4719}

\bibitem[\protect\citeauthoryear{{Govea}, {Gomez}, {Preston}  \&
  {Sneden}}{{Govea} et~al.}{2014}]{Govea2014}
{Govea} J.,  {Gomez} T.,  {Preston} G.~W.,   {Sneden} C.,  2014, \mn@doi [\apj]
  {10.1088/0004-637X/782/2/59}, \href
  {https://ui.adsabs.harvard.edu/abs/2014ApJ...782...59G} {782, 59}

\bibitem[\protect\citeauthoryear{{Graczyk} et~al.,}{{Graczyk}
  et~al.}{2020}]{Graczyk2020}
{Graczyk} D.,  et~al., 2020, \mn@doi [\apj] {10.3847/1538-4357/abbb2b}, \href
  {https://ui.adsabs.harvard.edu/abs/2020ApJ...904...13G} {904, 13}

\bibitem[\protect\citeauthoryear{{Gratton}, {Bragaglia}, {Clementini},
  {Carretta}, {Di Fabrizio}, {Maio}  \& {Taribello}}{{Gratton}
  et~al.}{2004}]{Gratton2004}
{Gratton} R.~G.,  {Bragaglia} A.,  {Clementini} G.,  {Carretta} E.,  {Di
  Fabrizio} L.,  {Maio} M.,   {Taribello} E.,  2004, \mn@doi [\aap]
  {10.1051/0004-6361:20035840}, \href
  {https://ui.adsabs.harvard.edu/abs/2004A&A...421..937G} {421, 937}

\bibitem[\protect\citeauthoryear{{Hajdu}, {D{\'e}k{\'a}ny}, {Catelan}, {Grebel}
   \& {Jurcsik}}{{Hajdu} et~al.}{2018}]{Hajdu2018}
{Hajdu} G.,  {D{\'e}k{\'a}ny} I.,  {Catelan} M.,  {Grebel} E.~K.,   {Jurcsik}
  J.,  2018, \mn@doi [\apj] {10.3847/1538-4357/aab4fd}, \href
  {https://ui.adsabs.harvard.edu/abs/2018ApJ...857...55H} {857, 55}

\bibitem[\protect\citeauthoryear{{Harris}}{{Harris}}{2010}]{Harris2010}
{Harris} W.~E.,  2010, \mn@doi [arXiv e-prints] {10.48550/arXiv.1012.3224},
  \href {https://ui.adsabs.harvard.edu/abs/2010arXiv1012.3224H} {p.
  arXiv:1012.3224}

\bibitem[\protect\citeauthoryear{{Hoffman} \& {Gelman}}{{Hoffman} \&
  {Gelman}}{2011}]{Hoffman2011}
{Hoffman} M.~D.,  {Gelman} A.,  2011, \mn@doi [arXiv e-prints]
  {10.48550/arXiv.1111.4246}, \href
  {https://ui.adsabs.harvard.edu/abs/2011arXiv1111.4246H} {p. arXiv:1111.4246}

\bibitem[\protect\citeauthoryear{{Huang} et~al.,}{{Huang}
  et~al.}{2021}]{Huang2021}
{Huang} Y.,  et~al., 2021, \mn@doi [\apj] {10.3847/1538-4357/abca37}, \href
  {https://ui.adsabs.harvard.edu/abs/2021ApJ...907...68H} {907, 68}

\bibitem[\protect\citeauthoryear{{Iorio} \& {Belokurov}}{{Iorio} \&
  {Belokurov}}{2019}]{Iorio2019}
{Iorio} G.,  {Belokurov} V.,  2019, \mn@doi [\mnras] {10.1093/mnras/sty2806},
  \href {https://ui.adsabs.harvard.edu/abs/2019MNRAS.482.3868I} {482, 3868}

\bibitem[\protect\citeauthoryear{{Iorio} \& {Belokurov}}{{Iorio} \&
  {Belokurov}}{2021}]{Iorio2021}
{Iorio} G.,  {Belokurov} V.,  2021, \mn@doi [\mnras] {10.1093/mnras/stab005},
  \href {https://ui.adsabs.harvard.edu/abs/2021MNRAS.502.5686I} {502, 5686}

\bibitem[\protect\citeauthoryear{{Jurcsik} \& {Kovacs}}{{Jurcsik} \&
  {Kovacs}}{1996}]{Jurcsik1996}
{Jurcsik} J.,  {Kovacs} G.,  1996, \aap, \href
  {https://ui.adsabs.harvard.edu/abs/1996A&A...312..111J} {312, 111}

\bibitem[\protect\citeauthoryear{{Lambert}, {Heath}, {Lemke}  \&
  {Drake}}{{Lambert} et~al.}{1996}]{Lambert1996}
{Lambert} D.~L.,  {Heath} J.~E.,  {Lemke} M.,   {Drake} J.,  1996, \mn@doi
  [\apjs] {10.1086/192274}, \href
  {https://ui.adsabs.harvard.edu/abs/1996ApJS..103..183L} {103, 183}

\bibitem[\protect\citeauthoryear{{Layden}}{{Layden}}{1994}]{Layden1994}
{Layden} A.~C.,  1994, \mn@doi [\aj] {10.1086/117132}, \href
  {https://ui.adsabs.harvard.edu/abs/1994AJ....108.1016L} {108, 1016}

\bibitem[\protect\citeauthoryear{{Li}, {Huang}, {Liu}, {Beers}  \&
  {Zhang}}{{Li} et~al.}{2023}]{Li2023}
{Li} X.-Y.,  {Huang} Y.,  {Liu} G.-C.,  {Beers} T.~C.,   {Zhang} H.-W.,  2023,
  \mn@doi [\apj] {10.3847/1538-4357/acadd5}, \href
  {https://ui.adsabs.harvard.edu/abs/2023ApJ...944...88L} {944, 88}

\bibitem[\protect\citeauthoryear{{Liu}, {Zhao}, {Chen}, {Takeda}  \&
  {Honda}}{{Liu} et~al.}{2013}]{Liu2013}
{Liu} S.,  {Zhao} G.,  {Chen} Y.-Q.,  {Takeda} Y.,   {Honda} S.,  2013, \mn@doi
  [Research in Astronomy and Astrophysics] {10.1088/1674-4527/13/11/003}, \href
  {https://ui.adsabs.harvard.edu/abs/2013RAA....13.1307L} {13, 1307}

\bibitem[\protect\citeauthoryear{{Liu} et~al.,}{{Liu} et~al.}{2020}]{Liu2020}
{Liu} G.~C.,  et~al., 2020, \mn@doi [\apjs] {10.3847/1538-4365/ab72f8}, \href
  {https://ui.adsabs.harvard.edu/abs/2020ApJS..247...68L} {247, 68}

\bibitem[\protect\citeauthoryear{{Longmore}, {Fernley}  \&
  {Jameson}}{{Longmore} et~al.}{1986}]{Longmore1986}
{Longmore} A.~J.,  {Fernley} J.~A.,   {Jameson} R.~F.,  1986, \mn@doi [\mnras]
  {10.1093/mnras/220.2.279}, \href
  {https://ui.adsabs.harvard.edu/abs/1986MNRAS.220..279L} {220, 279}

\bibitem[\protect\citeauthoryear{{Madore} et~al.,}{{Madore}
  et~al.}{2013}]{Madore2013}
{Madore} B.~F.,  et~al., 2013, \mn@doi [\apj] {10.1088/0004-637X/776/2/135},
  \href {https://ui.adsabs.harvard.edu/abs/2013ApJ...776..135M} {776, 135}

\bibitem[\protect\citeauthoryear{{Moln{\'a}r}, {P{\'a}l}, {Plachy}, {Ripepi},
  {Moretti}, {Szab{\'o}}  \& {Kiss}}{{Moln{\'a}r} et~al.}{2015}]{Molnar2015}
{Moln{\'a}r} L.,  {P{\'a}l} A.,  {Plachy} E.,  {Ripepi} V.,  {Moretti} M.~I.,
  {Szab{\'o}} R.,   {Kiss} L.~L.,  2015, \mn@doi [\apj]
  {10.1088/0004-637X/812/1/2}, \href
  {https://ui.adsabs.harvard.edu/abs/2015ApJ...812....2M} {812, 2}

\bibitem[\protect\citeauthoryear{{Morgan}, {Wahl}  \& {Wieckhorst}}{{Morgan}
  et~al.}{2007}]{Morgan2007}
{Morgan} S.~M.,  {Wahl} J.~N.,   {Wieckhorst} R.~M.,  2007, \mn@doi [\mnras]
  {10.1111/j.1365-2966.2006.11247.x}, \href
  {https://ui.adsabs.harvard.edu/abs/2007MNRAS.374.1421M} {374, 1421}

\bibitem[\protect\citeauthoryear{{Mullen} et~al.,}{{Mullen}
  et~al.}{2021}]{Mullen2021}
{Mullen} J.~P.,  et~al., 2021, \mn@doi [\apj] {10.3847/1538-4357/abefd4}, \href
  {https://ui.adsabs.harvard.edu/abs/2021ApJ...912..144M} {912, 144}

\bibitem[\protect\citeauthoryear{{Muraveva} et~al.,}{{Muraveva}
  et~al.}{2018a}]{Muraveva2018c}
{Muraveva} T.,  et~al., 2018a, \mn@doi [\mnras] {10.1093/mnras/stx2514}, \href
  {https://ui.adsabs.harvard.edu/abs/2018MNRAS.473.3131M} {473, 3131}

\bibitem[\protect\citeauthoryear{{Muraveva}, {Garofalo}, {Scowcroft},
  {Clementini}, {Freedman}, {Madore}  \& {Monson}}{{Muraveva}
  et~al.}{2018b}]{Muraveva2018a}
{Muraveva} T.,  {Garofalo} A.,  {Scowcroft} V.,  {Clementini} G.,  {Freedman}
  W.~L.,  {Madore} B.~F.,   {Monson} A.~J.,  2018b, \mn@doi [\mnras]
  {10.1093/mnras/sty1959}, \href
  {https://ui.adsabs.harvard.edu/abs/2018MNRAS.480.4138M} {480, 4138}

\bibitem[\protect\citeauthoryear{{Muraveva}, {Delgado}, {Clementini}, {Sarro}
  \& {Garofalo}}{{Muraveva} et~al.}{2018c}]{Muraveva2018b}
{Muraveva} T.,  {Delgado} H.~E.,  {Clementini} G.,  {Sarro} L.~M.,   {Garofalo}
  A.,  2018c, \mn@doi [\mnras] {10.1093/mnras/sty2241}, \href
  {https://ui.adsabs.harvard.edu/abs/2018MNRAS.481.1195M} {481, 1195}

\bibitem[\protect\citeauthoryear{{Muraveva}, {Clementini}, {Garofalo}  \&
  {Cusano}}{{Muraveva} et~al.}{2020}]{Muraveva2020}
{Muraveva} T.,  {Clementini} G.,  {Garofalo} A.,   {Cusano} F.,  2020, \mn@doi
  [\mnras] {10.1093/mnras/staa2984}, \href
  {https://ui.adsabs.harvard.edu/abs/2020MNRAS.499.4040M} {499, 4040}

\bibitem[\protect\citeauthoryear{{Nemec}, {Cohen}, {Ripepi}, {Derekas},
  {Moskalik}, {Sesar}, {Chadid}  \& {Bruntt}}{{Nemec} et~al.}{2013}]{Nemec2013}
{Nemec} J.~M.,  {Cohen} J.~G.,  {Ripepi} V.,  {Derekas} A.,  {Moskalik} P.,
  {Sesar} B.,  {Chadid} M.,   {Bruntt} H.,  2013, \mn@doi [\apj]
  {10.1088/0004-637X/773/2/181}, \href
  {https://ui.adsabs.harvard.edu/abs/2013ApJ...773..181N} {773, 181}

\bibitem[\protect\citeauthoryear{{Pancino}, {Britavskiy}, {Romano}, {Cacciari},
  {Mucciarelli}  \& {Clementini}}{{Pancino} et~al.}{2015}]{Pancino2015}
{Pancino} E.,  {Britavskiy} N.,  {Romano} D.,  {Cacciari} C.,  {Mucciarelli}
  A.,   {Clementini} G.,  2015, \mn@doi [\mnras] {10.1093/mnras/stu2616}, \href
  {https://ui.adsabs.harvard.edu/abs/2015MNRAS.447.2404P} {447, 2404}

\bibitem[\protect\citeauthoryear{Pedregosa et~al.,}{Pedregosa
  et~al.}{2011}]{scikit-learn}
Pedregosa F.,  et~al., 2011, Journal of Machine Learning Research, 12, 2825

\bibitem[\protect\citeauthoryear{{Pietrzy{\'n}ski} et~al.,}{{Pietrzy{\'n}ski}
  et~al.}{2019}]{Pietrzynski2019}
{Pietrzy{\'n}ski} G.,  et~al., 2019, \mn@doi [\nat]
  {10.1038/s41586-019-0999-4}, \href
  {https://ui.adsabs.harvard.edu/abs/2019Natur.567..200P} {567, 200}

\bibitem[\protect\citeauthoryear{{Plachy} et~al.,}{{Plachy}
  et~al.}{2021}]{Plachy2021}
{Plachy} E.,  et~al., 2021, \mn@doi [\apjs] {10.3847/1538-4365/abd4e3}, \href
  {https://ui.adsabs.harvard.edu/abs/2021ApJS..253...11P} {253, 11}

\bibitem[\protect\citeauthoryear{{Preston}}{{Preston}}{1959}]{Preston1959}
{Preston} G.~W.,  1959, \mn@doi [\apj] {10.1086/146743}, \href
  {https://ui.adsabs.harvard.edu/abs/1959ApJ...130..507P} {130, 507}

\bibitem[\protect\citeauthoryear{{Schlegel}, {Finkbeiner}  \&
  {Davis}}{{Schlegel} et~al.}{1998}]{sfd1998}
{Schlegel} D.~J.,  {Finkbeiner} D.~P.,   {Davis} M.,  1998, \mn@doi [\apj]
  {10.1086/305772}, \href
  {https://ui.adsabs.harvard.edu/abs/1998ApJ...500..525S} {500, 525}

\bibitem[\protect\citeauthoryear{{Scowcroft}, {Freedman}, {Madore}, {Monson},
  {Persson}, {Rich}, {Seibert}  \& {Rigby}}{{Scowcroft}
  et~al.}{2016}]{Skowcroft2016}
{Scowcroft} V.,  {Freedman} W.~L.,  {Madore} B.~F.,  {Monson} A.,  {Persson}
  S.~E.,  {Rich} J.,  {Seibert} M.,   {Rigby} J.~R.,  2016, \mn@doi [\apj]
  {10.3847/0004-637X/816/2/49}, \href
  {https://ui.adsabs.harvard.edu/abs/2016ApJ...816...49S} {816, 49}

\bibitem[\protect\citeauthoryear{{Sesar} et~al.,}{{Sesar}
  et~al.}{2014}]{Sesar2014}
{Sesar} B.,  et~al., 2014, \mn@doi [\apj] {10.1088/0004-637X/793/2/135}, \href
  {https://ui.adsabs.harvard.edu/abs/2014ApJ...793..135S} {793, 135}

\bibitem[\protect\citeauthoryear{{Skowron} et~al.,}{{Skowron}
  et~al.}{2016}]{Skowron2016}
{Skowron} D.~M.,  et~al., 2016, \mn@doi [\actaa] {10.48550/arXiv.1608.00013},
  \href {https://ui.adsabs.harvard.edu/abs/2016AcA....66..269S} {66, 269}

\bibitem[\protect\citeauthoryear{{Skowron} et~al.,}{{Skowron}
  et~al.}{2021}]{Skowron2021}
{Skowron} D.~M.,  et~al., 2021, \mn@doi [\apjs] {10.3847/1538-4365/abcb81},
  \href {https://ui.adsabs.harvard.edu/abs/2021ApJS..252...23S} {252, 23}

\bibitem[\protect\citeauthoryear{{Smolec}}{{Smolec}}{2005}]{Smolec2005}
{Smolec} R.,  2005, \mn@doi [\actaa] {10.48550/arXiv.astro-ph/0503614}, \href
  {https://ui.adsabs.harvard.edu/abs/2005AcA....55...59S} {55, 59}

\bibitem[\protect\citeauthoryear{{Sneden}, {Preston}, {Chadid}  \&
  {Adam{\'o}w}}{{Sneden} et~al.}{2017}]{Sneden2017}
{Sneden} C.,  {Preston} G.~W.,  {Chadid} M.,   {Adam{\'o}w} M.,  2017, \mn@doi
  [\apj] {10.3847/1538-4357/aa8b10}, \href
  {https://ui.adsabs.harvard.edu/abs/2017ApJ...848...68S} {848, 68}

\bibitem[\protect\citeauthoryear{{Sollima}, {Cacciari}, {Arkharov}, {Larionov},
  {Gorshanov}, {Efimova}  \& {Piersimoni}}{{Sollima}
  et~al.}{2008}]{Sollima2008}
{Sollima} A.,  {Cacciari} C.,  {Arkharov} A.~A.~H.,  {Larionov} V.~M.,
  {Gorshanov} D.~L.,  {Efimova} N.~V.,   {Piersimoni} A.,  2008, \mn@doi
  [\mnras] {10.1111/j.1365-2966.2007.12804.x}, \href
  {https://ui.adsabs.harvard.edu/abs/2008MNRAS.384.1583S} {384, 1583}

\bibitem[\protect\citeauthoryear{{Soszy{\'n}ski} et~al.,}{{Soszy{\'n}ski}
  et~al.}{2014}]{OGLE1}
{Soszy{\'n}ski} I.,  et~al., 2014, \mn@doi [\actaa] {10.48550/arXiv.1410.1542},
  \href {https://ui.adsabs.harvard.edu/abs/2014AcA....64..177S} {64, 177}

\bibitem[\protect\citeauthoryear{{Soszy{\'n}ski} et~al.,}{{Soszy{\'n}ski}
  et~al.}{2019}]{OGLE2}
{Soszy{\'n}ski} I.,  et~al., 2019, \mn@doi [\actaa]
  {10.32023/0001-5237/69.4.2}, \href
  {https://ui.adsabs.harvard.edu/abs/2019AcA....69..321S} {69, 321}

\bibitem[\protect\citeauthoryear{{Zinn} \& {West}}{{Zinn} \&
  {West}}{1984}]{ZW1984}
{Zinn} R.,  {West} M.~J.,  1984, \mn@doi [\apjs] {10.1086/190947}, \href
  {https://ui.adsabs.harvard.edu/abs/1984ApJS...55...45Z} {55, 45}

\bibitem[\protect\citeauthoryear{{Zinn}, {Chen}, {Layden}  \&
  {Casetti-Dinescu}}{{Zinn} et~al.}{2020}]{Zinn2020}
{Zinn} R.,  {Chen} X.,  {Layden} A.~C.,   {Casetti-Dinescu} D.~I.,  2020,
  \mn@doi [\mnras] {10.1093/mnras/stz3580}, \href
  {https://ui.adsabs.harvard.edu/abs/2020MNRAS.492.2161Z} {492, 2161}

\makeatother
\end{thebibliography}




\appendix

\section{Dataset}

\begin{landscape} 
\begin{table} 
\caption{Parameters of the 150 RRLs in HR-CAT-RRLS sample: (1) {\it Gaia} DR3 source\_id; (2) and (3) Coordinates; (4) RRL type; (5) Period; (6) Intensity-averaged $G$ magnitude; (7)-(11) Fourier parameters; (12) Metallicity from HR-spectroscopy on the metallicity scale adopted by \citet{Crestani2021}; (13) Source of metallicity (1 - \citealt{Crestani2021}, 2 - \citealt{For2011}, 3 -\citealt{Sneden2017}, 4 - \citealt{Chadid2017}, 5 - \citealt{Gilligan2021}).
Columns (1)-(3) are from {\it Gaia} DR3 \texttt{gaia\_source} table \citep{Vallenari2023}. Columns (4)-(11) are from {\it Gaia} DR3 \texttt{vari\_rrlyrae} table \citep{Clementini2023}.}
\label{tab:cat_hr} 
\begin{tabular}{ccccccccccccccccccc} 
\hline 
source\_id & RA & DEC & Type & Period & $G$ &  $A_1$ & $A_2$ & $A_3$ & $\phi_{21}$ & $\phi_{31}$ & [Fe/H] & Source \\
\hline
  77849374617106176   & 30.1396  &14.1983 & RRab      &   0.5576 & 15.67& $0.405\pm 0.036$ &   $0.167 \pm0.045$  &  $ -        $      &  $3.55 \pm 0.12$ &     $   -       $     &     $-1.78 \pm0.09$  &    1    \\ 
  630421935431871232  & 151.9310 &23.9917 & RRab      &          0.4524 & 10.65& $0.405\pm 0.014$ &   $0.177 \pm0.008$  &  $0.121 \pm0.012 $ &  $3.80 \pm 0.08$ &     $1.64 \pm 0.11 $  &     $-1.58 \pm0.27$  &    1    \\ 
  1191510003353849472 & 238.3794 &12.9611 & RRab      &          0.5221 & 10.84& $0.322\pm 0.005$ &   $0.160 \pm0.009$  &  $0.085 \pm0.010 $ &  $4.32 \pm 0.05$ &     $2.59 \pm 0.05 $  &     $ 0.05 \pm0.15$  &    4    \\ 
  1234729400256865664 & 221.8968 &16.8453 & RRc       &          0.3149 & 10.58& $0.182\pm 0.004$ &   $0.016 \pm0.006$  &  $0.014 \pm0.005 $ &  $4.52 \pm 0.28$ &     $4.08 \pm 0.25 $  &     $-1.62 \pm0.15$  &    1    \\
  1453674738379109760 & 209.3918 &29.8579 & RRc       &          0.3290 & 11.29& $0.181\pm 0.003$ &   $0.014 \pm0.004$  &  $0.013 \pm0.005 $ &  $4.82 \pm 0.32$ &     $4.45 \pm 0.25 $  &     $-1.46 \pm0.15$  &    1    \\
  1492230556717187456 & 214.1524 &42.3598 & RRc       &          0.3126 & 10.91& $0.238\pm 0.005$ &   $0.075 \pm0.005$  &  $0.020 \pm0.004 $ &  $4.32 \pm 0.06$ &     $2.37 \pm 0.22 $  &     $-2.54 \pm0.13$  &    1    \\ 
  1565435491138161664 & 201.5555 &56.2570 & RRc       &          0.3071 & 10.79& $0.231\pm 0.003$ &   $0.050 \pm0.003$  &  $0.014 \pm0.003 $ &  $4.66 \pm 0.06$ &     $2.94 \pm 0.25 $  &     $-1.73 \pm0.15$  &    1    \\
  1760981190300823808 & 311.8682 &12.4641 & RRab      &          0.4726 & 9.81 & $0.227\pm 0.002$ &   $0.112 \pm0.002$  &  $0.065 \pm0.002 $ &  $4.27 \pm 0.02$ &     $2.51 \pm 0.04 $  &     $-0.43 \pm0.17$  &    4    \\
  1770039418063209600 & 323.2143 &12.8919 & RRab      &          0.6005 & 13.91& $0.232\pm 0.009$ &   $0.101 \pm0.010$  &  $0.064 \pm0.008 $ &  $4.02 \pm 0.06$ &     $1.97 \pm 0.16 $  &     $-1.68 \pm0.11$  &    1    \\ 
  1786827307055763968 & 321.7645 &18.5992 & RRab      &          0.5472 & 12.77& $0.287\pm 0.015$ &   $0.140 \pm0.011$  &  $0.097 \pm0.011 $ &  $3.91 \pm 0.09$ &     $1.92 \pm 0.13 $  &     $-1.67 \pm0.12$  &    1    \\ 
\hline

\end{tabular} 

The table is published in its entirety as Supporting Information with the electronic version of the article. A portion is shown here for guidance regarding its form and content.

\end{table} 
\end{landscape}

\begin{landscape} 
\begin{table} 
\caption{Parameters of 134,769 RRLs, for which photometric metallicities were calculated in this study: (1) {\it Gaia} DR3 source\_id; (2) and (3) Coordinates; (4) RRL type; (5) Period; (6)-(10) Fourier parameters; (11) Photometric metallicity on the scale adopted by \citet{Crestani2021}; (12) Reddening values compiled as described in Section~\ref{sec:dist}; (13) Distance modulus. Columns (1)-(3) are from {\it Gaia} DR3 \texttt{gaia\_source} table \citep{Vallenari2023}. Columns (4)-(10) are from {\it Gaia} DR3 \texttt{vari\_rrlyrae} table \citep{Clementini2023}.}\label{tab:cat_gaia} 
\begin{tabular}{ccccccccccccc} 

\hline 
 
source\_id & RA & DEC & Type & Period & $A_1$ & $A_2$ & $A_3$ & $\phi_{21}$ & $\phi_{31}$ & [Fe/H] & $E(B-V)$ & $\mu$ \\
\hline
  500243431117184   &  44.6981 &  1.7456  & RRab        &          0.6231 & $0.311\pm 0.019$ & $0.155\pm 0.026$    & $0.095\pm 0.019$ &  $3.95 \pm  0.13 $  &      $1.89\pm   0.32$   &      $-2.05\pm 0.50$  &    $ 0.090\pm 0.002 $   &      $ 17.53 \pm 0.99 $    \\         
  507222753405440   &  44.6653 &  1.7894  & RRab        &          0.6089 & $0.335\pm 0.033$ & $0.190\pm 0.030$    & $0.129\pm 0.020$ &  $4.03 \pm  0.13 $  &      $1.93\pm   0.31$   &      $-1.93\pm 0.48$  &    $ 0.092\pm 0.003 $   &      $ 14.02 \pm 0.95 $    \\        
  584630948352256   &  46.3415 &  1.5420  & RRab        &          0.5555 & $0.278\pm 0.020$ & $0.137\pm 0.015$    & $0.080\pm 0.018$ &  $3.96 \pm  0.13 $  &      $2.31\pm   0.22$   &      $-1.28\pm 0.44$  &    $ 0.077\pm 0.001 $   &      $ 16.61 \pm 0.76 $    \\         
  782027645388032   &  46.6991 &  2.3039  & RRab        &          0.5583 & $0.277\pm 0.028$ & $0.154\pm 0.034$    & $0.108\pm 0.038$ &  $4.43 \pm  0.30 $  &      $2.38\pm   0.44$   &      $-1.23\pm 0.57$  &    $ 0.087\pm 0.009 $   &      $ 17.09 \pm 0.77 $    \\         
  1407379179248512  &  42.8875 &  1.8053  & RRab        &          0.5813 & $0.257\pm 0.017$ & $0.116\pm 0.015$    & $0.105\pm 0.017$ &  $4.07 \pm  0.16 $  &      $1.93\pm   0.21$   &      $-1.78\pm 0.44$  &    $ 0.058\pm 0.001 $   &      $ 16.90 \pm 0.91 $    \\         
  1514169246023424  &  42.8608 &  2.4539  & RRab        &          0.6144 & $0.295\pm 0.073$ & $0.120\pm 0.087$    & $0.116\pm 0.075$ &  $4.15 \pm  0.32 $  &      $1.87\pm   0.49$   &      $-2.02\pm 0.63$  &    $ 0.087\pm 0.001 $   &      $ 15.13 \pm 1.00 $    \\        
  1742622850845696  &  45.2336 &  2.9300  & RRab        &          0.4565 & $0.342\pm 0.022$ & $0.148\pm 0.032$    & $0.078\pm 0.020$ &  $3.73 \pm  0.22 $  &      $1.87\pm   0.33$   &      $-1.15\pm 0.48$  &    $ 0.091\pm 0.003 $   &      $ 17.22 \pm 0.74 $    \\         
  2345017784412032  &  47.2341 &  2.9847  & RRab        &          0.6143 & $0.237\pm 0.013$ & $0.127\pm 0.019$    & $0.087\pm 0.018$ &  $4.14 \pm  0.14 $  &      $2.41\pm   0.43$   &      $-1.51\pm 0.56$  &    $ 0.124\pm 0.001 $   &      $ 16.99 \pm 0.84 $    \\         
  3070042623687040  &  50.1084 &  4.9737  & RRab        &          0.5974 & $0.220\pm 0.008$ & $0.094\pm 0.009$    & $0.058\pm 0.011$ &  $3.97 \pm  0.12 $  &      $2.12\pm   0.20$   &      $-1.70\pm 0.42$  &    $ 0.160\pm 0.019 $   &      $ 17.48 \pm 0.87 $    \\

\hline
\end{tabular} 

The table is published in its entirety as Supporting Information with the electronic version of the article. A portion is shown here for guidance regarding its form and content.

\end{table} 
\end{landscape}

\section{XGBoost analysis}\label{sec:xgboost}

Our Bayesian analysis of the connection between light curve parameters and the metallicity of RRLs assumes a linear correlation, a simplifying assumption that may not capture more intricate, subtle dependencies within the data. To explore the potential existence of complex and weak relationships, we employed an XGBoost regressor to predict the metallicity using the {\it Gaia} DR3 data, with the RRab stars in the HR-CAT-RRLS catalogue serving as the training set.

The decision to utilize the XGBoost algorithm was driven by the many advantages it provides for the task we set out to accomplish.
The efficiency of XGBoost in handling non-linear relationships, together with its adeptness at handling complex tabular data played a pivotal role in our choice.
Another compelling factor in favor of XGBoost is its capacity to deliver exceptionally high performance and robustness, even when confronted with relatively small datasets.  XGBoost's resilience makes it a reliable choice for extracting meaningful insights from the limited observational data available.

Furthermore, the time efficiency of XGBoost is noteworthy. Its capability to complete full training experiments within a limited time frame is advantageous, allowing to allocate more time to fine-tuning the algorithm and augmenting the dataset. This not only enhances the model's predictive accuracy but also provides an opportunity for a more comprehensive exploration of the parameter space.

Our dataset was randomly split into training and validation sets, with the training set representing $80\%$ of the full catalogue, with the rest being held out as validation set.
Before starting the training process, we applied a quantile transformation with uniform output to the data, using all the features of the light curve Fourier decomposition and period. After scaling the data, we expanded the dataset by creating additional columns that quantify the similarity of each input to the average values within specific groups. These groups were defined by segregating the data into bins based on their metallicity values. Each column in this series reflects the degree of resemblance between an input and the average values associated with its respective metallicity bin.
In practice, we divided the dataset in ten subsamples based on the sample value in terms of metallicity and computed the average values of the input features per bin. The bins were chosen so that no bin was empty. Then, for each sample, we computed the cosine similarity of the input features to the mean values per metallicity bin, including the result in ten additional feature columns, one per bin, following the idea that similar inputs should lead to similar outputs.
During training we also applied sample weighting. The weights correspond to the inverse of the KDE smoothing of the metallicity computed on the training set, with a Gaussian kernel and a bandwidth of $0.2$.

Performances were optimized by the GridSearchCV function of the SKLearn package \citep{scikit-learn}. We used a threefold cross validation, and optimized the number of estimators, the maximum depth of the trees, the learning rate, the lamdba regularization, the subsampling of the training set and the columns used per tree and per level. The best RMSE reached on the validation set is $0.29$~dex. The corresponding hyperparameters are given in Table~\ref{tab:xgb_parameters}.

\begin{table}
  \centering
  \begin{tabular}{lr}
    \hline
    \hline
    \textbf{Parameter} & \textbf{Value} \\
    \hline
    n\_estimators & 400 \\
    max\_depth & 3 \\
    learning\_rate & 0.09 \\
    lambda & 1 \\
    subsample & 0.6 \\
    colsample\_bylevel & 0.7 \\
    colsample\_bytree & 0.9 \\
    \hline
    \hline
  \end{tabular}
  \caption{Hyperparameters of the XGBoost regressor with the best performances reached.}
  \label{tab:xgb_parameters}
\end{table}

We repeated all test that were carried out for the Bayesian approach, but we did not find a significant increase in performances, possibly indicating that the relationships between the light curve Fourier parameters, periods and the metallicity are indeed linear. It could also be that the data at our disposal are not sufficient to detect more complex correlations, on the account of the training set being too small. In fact, inspecting the learning curve of the model reveals that the RMSE of the training set flattens to values close to the intrinsic scatter of the relation that we find with our Bayesian approach, while that in the validation set flattens to the higher values reported above, suggesting that adding more data may allow the model to refine the learned relations.


\bsp	
\label{lastpage}
\end{document}